\documentclass[prx,twocolumn,amsmath,amssymb,superscriptaddress]{revtex4-2}

\usepackage{natbib}
\usepackage{amsmath} % AMS Math Package
\usepackage{amssymb} % Math symbols such as \mathbb
\usepackage{graphicx} % Allows for eps images
\usepackage{times}
\usepackage{color}
\usepackage{booktabs}
\usepackage{siunitx}

\usepackage{ulem}

\begin{document}

\title{Spin-orbit coupling in a half-filled $t_{2g}$ shell: the case of $5d^3$ K$_2$ReCl$_6$}

\author{P. Warzanowski}
\author{M. Magnaterra}
\author{G. Schlicht}
\affiliation{Institute of Physics II, University of Cologne, 50937 Cologne, Germany}
\author{Q. Faure}
\affiliation{ESRF, The European Synchrotron, 71 Avenue des Martyrs, CS40220, 38043 Grenoble Cedex 9, France}
\affiliation{Laboratoire L\'{e}on Brillouin, CEA, CNRS, Universit\'{e} Paris-Saclay, CEA-Saclay, 
	91191 Gif-sur-Yvette, France}
\author{Ch. J. Sahle}
\affiliation{ESRF, The European Synchrotron, 71 Avenue des Martyrs, CS40220, 38043 Grenoble Cedex 9, France}
\author{P. Becker}
\author{L.~Bohat\'{y}}
\affiliation{Sect. Crystallography, Institute of Geology and Mineralogy, University of Cologne, 50674 Cologne, Germany}
\author{M. Moretti Sala}
\affiliation{Dipartimento di Fisica, Politecnico di Milano, I-20133 Milano, Italy}
\author{G. Monaco}
\affiliation{Dipartimento di Fisica e Astronomia "Galileo Galilei", Universit\`{a} di Padova, I-35121 Padova, Italy}
\author{M. Hermanns}
\affiliation{Department of Physics, Stockholm University, AlbaNova University Center, SE-106 91 Stockholm, Sweden}
\author{P.H.M. van Loosdrecht}
\author{M. Gr\"{u}ninger}
\affiliation{Institute of Physics II, University of Cologne, 50937 Cologne, Germany}

\begin{abstract}
The half-filled $t_{2g}$ shell of the $t_{2g}^3$ configuration usually, in $LS$ coupling, 
hosts a $S$\,=\,3/2 ground state with quenched orbital moment. This state is not Jahn-Teller 
active. Sufficiently large spin-orbit coupling $\zeta$ has been predicted to change this picture 
by mixing in orbital moment, giving rise to a sizable Jahn-Teller distortion. 
In $5d^3$ K$_2$ReCl$_6$ we study the electronic excitations using resonant inelastic 
x-ray scattering (RIXS) and optical spectroscopy. We observe on-site intra-$t_{2g}$ 
excitations below 2\,eV and corresponding overtones with two intra-$t_{2g}$ excitations 
on adjacent sites, the Mott gap at 2.7\,eV, $t_{2g}$-to-$e_g$ excitations above 3\,eV, 
and charge-transfer excitations at still higher energy. 
The intra-$t_{2g}$ excitation energies are a sensitive measure of $\zeta$ and Hund's 
coupling $J_{\rm H}$. The sizable value of $\zeta$\,$\approx$\,0.29\,eV places 
K$_2$ReCl$_6$ into the intermediate coupling regime, but 
$\zeta/J_{\rm H}$\,$\approx$\,0.6 is not sufficiently large 
to drive a pronounced Jahn-Teller effect. We discuss the ground state wavefunction in a 
Kanamori picture and find that the $S$\,=\,3/2 multiplet still carries about 97\,\% of the weight. 
However, the finite admixture of orbital moment allows for subtle effects. 
We discuss small temperature-induced changes of the optical data and find evidence for a lowering 
of the ground state by about 3\,meV below the structural phase transitions. 
\end{abstract}

\date{November 19, 2023}

\maketitle

\section{Introduction}

Strong spin-orbit coupling $\zeta$ is a fertile source of rich physics in correlated $5d$ 
transition-metal compounds \cite{WitczakKrempa14,Rau16,Schaffer16,Streltsov20,Takayama21,Khomskii21}. 
Consider a single metal site with the electronic configuration $t_{2g}^n$ in cubic symmetry. 
For $n$\,=\,1, 2, 4, or 5 and $\zeta$\,=\,0, the ground state shows either spin $S$\,=\,1/2 
or 1 and threefold orbital degeneracy, and the latter is expected to be lifted by the 
Jahn-Teller effect \cite{Streltsov20}. 
In all four cases, strong spin-orbit coupling changes the character of the magnetic moments 
in a decisive way, forming spin-orbit-entangled moments from the spin and the effective orbital 
moment of the $t_{2g}$ states.  
The $t_{2g}^5$ configuration hosts $J$\,=\,1/2 moments that open the door for the realization 
of bond-directional Kitaev exchange \cite{Jackeli09,Winter17,Chun15,Revelli19a,Magnaterra23}. 
Similarly, $n$\,=\,1 yields $J$\,=\,3/2 moments with bond-dependent multipolar interactions 
and a corresponding rich phase diagram 
\cite{Chen10,Natori16,Romhanyi17,Ishikawa19,Tehrani23}. 
Exotic multipolar phases have also been predicted for $t_{2g}^2$ compounds where the 
degeneracy of the $J$\,=\,2 states is lifted if one considers the admixture of $e_g$ orbitals 
for a finite cubic crystal-field splitting 10\,$Dq$ \cite{Paramekanti20,Lovesey20,Khaliullin21,Voleti21,Pourovskii21,Rayyan23}. 
Finally, $t_{2g}^4$ compounds show a non-magnetic $J$\,=\,0 state 
\cite{Warzanowski23,Yuan17,Kusch18,Nag18,Aczel22,Paramekanti18,Fuchs18,Takahashi21} 
that, in the case of strong dispersion of magnetic excited states, may give way to excitonic 
magnetism based on the condensation of these excited states 
\cite{Khaliullin13,KhomskiiBook,Jain17,Kaushal21}.

\begin{figure}[t]
	\centering
	\includegraphics[width=0.91\columnwidth]{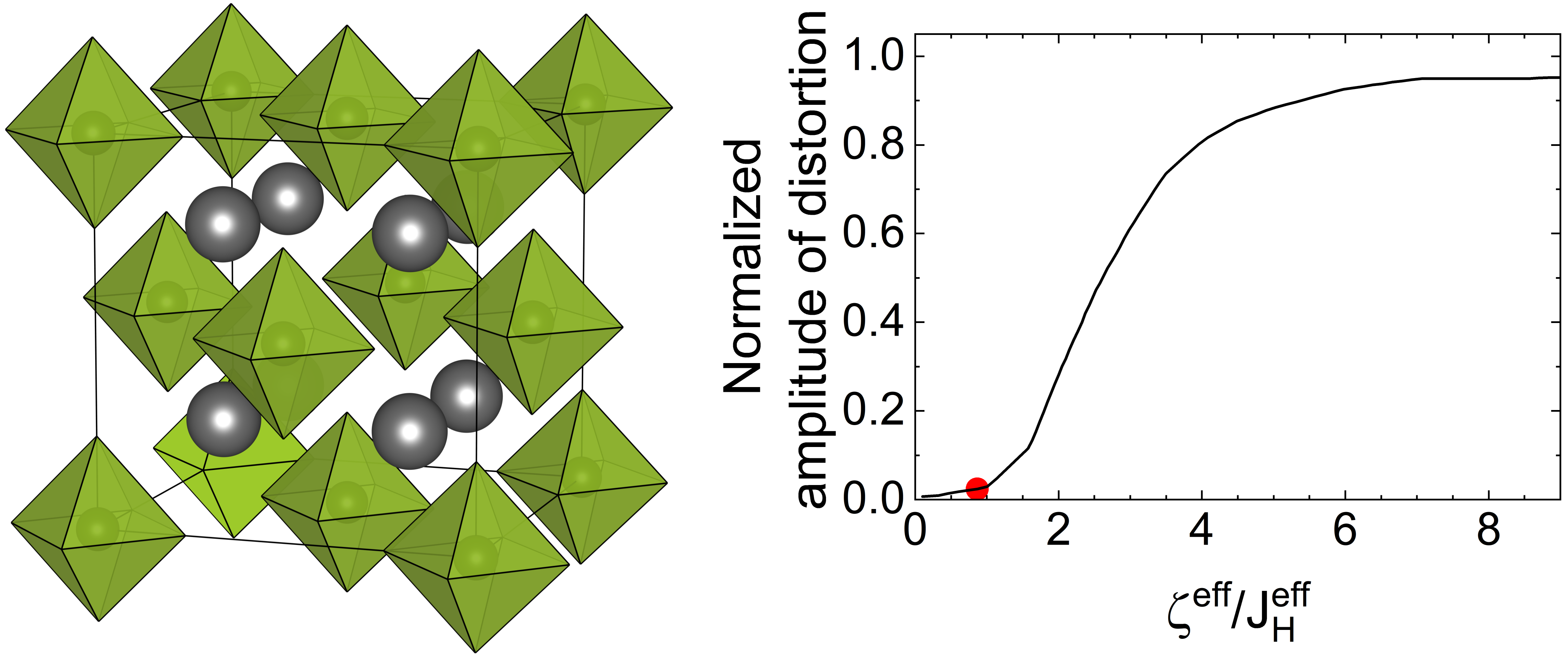}
	\caption{Left: Sketch of the room-temperature crystal structure of K$_2$ReCl$_6$. 
	The Re$^{4+}$ ions form an \textit{fcc} lattice. 
	The ReCl$_6$ octahedra are depicted in green and the K$^+$ ions are gray. 
	Right: Jahn-Teller distortion of a $t_{2g}^3$ configuration as a function of 
	$\zeta^{\rm eff}/J_{\rm H}^{\rm eff}$, normalized to the value for 
	$\zeta^{\rm eff}/J_{\rm H}^{\rm eff}$\,$\rightarrow$\,$\infty$ (adapted from \cite{Streltsov20}). 
	The calculation assumes $J_{\rm H} B/g^2$\,=\,1/2, where $g$ and $B$ are electron-phonon 
	coupling parameters of the Jahn-Teller Hamiltonian \cite{Streltsov20}. 
	Red symbol marks the position of K$_2$ReCl$_6$ according to our analysis. 
	}
	\label{fig:structure}
\end{figure}

In this series, the case of $n$\,=\,3 stands out due to its half-filled $t_{2g}$ shell, which 
typically is assumed to give rise to a spin-only $S$\,=\,3/2 state with quenched orbital moment. 
Spin-orbit coupling hence is not expected to play a prominent role, at least in the commonly 
adopted $LS$ coupling scheme for $\zeta/J_{\rm H}$\,$\rightarrow$\,0, where 
$J_{\rm H}$ denotes Hund's coupling.
Recently, Streltsov and Khomskii \cite{Streltsov20} pointed out that this common point of view 
fails for large $\zeta/J_{\rm H}$, highlighting a particularly interesting case of the interplay of 
spin-orbit coupling and Jahn-Teller physics. In the $S$\,=\,3/2 scenario, the three electrons 
equally occupy the three $t_{2g}$ orbitals such that a distortion away from cubic symmetry 
does not lower the energy. In contrast, for large $\zeta/J_{\rm H}$ in $jj$ coupling one obtains 
$j$\,=\,3/2 for each electron individually, and the corresponding 3-electron ground state 
is found to be Jahn-Teller active \cite{Streltsov20}. More precisely, a significant Jahn-Teller 
distortion is expected for $\zeta^{\rm eff}/J_{\rm H}^{\rm eff} \gtrsim 1.5$, see Fig.\ \ref{fig:structure}b). 
The effective parameters $\zeta^{\rm eff}$ and $J_{\rm H}^{\rm eff}$ refer to a $t_{2g}$-only 
Kanamori scheme, i.e., to the case of an infinite cubic crystal-field splitting, 
10\,$Dq$\,=\,$\infty$.

Experimentally, the electronic parameters $\zeta$ and $J_{\rm H}$ can be determined via the 
energies of intra-$t_{2g}$ excitations that can be observed in resonant inelastic x-ray 
scattering (RIXS) or optical spectroscopy. 
Based on RIXS data, the $5d^3$ osmates Ca$_3$LiOsO$_6$ and Ba$_2$YOsO$_6$ have been claimed 
to realize a novel spin-orbit-entangled $J$\,=\,3/2 ground state for which, however, about 
95\,\% of the wavefunction stem from the $S$\,=\,3/2 state \cite{Taylor17}. The admixture of 
low-spin $S$\,=\,1/2 character and the corresponding finite orbital moment are supposed 
to explain the sizable spin gap observed in both 
compounds and other $5d^3$ osmates \cite{Kermarec15,Calder16,Taylor16,Taylor18}. 
In contrast, gapless magnetic excitations were reported for $4d^3$ $S$\,=\,3/2 
Ca$_3$LiRuO$_6$ \cite{Calder17b}. 
In the RIXS data of $5d^3$ Ca$_3$LiOsO$_6$ and Ba$_2$YOsO$_6$, only four of the five 
intra-$t_{2g}$ excitations were resolved \cite{Taylor17}. The analysis yields 
$\zeta/J_{\rm H}$\,$\approx$\,1, placing these compounds within the intermediate 
coupling range \cite{Taylor17}.

Here, we address the electronic structure of the $5d^3$ Mott insulator K$_2$ReCl$_6$ using  
RIXS measurements at the Re $L_3$ edge and optical spectroscopy. At room temperature, 
K$_2$ReCl$_6$ exhibits the cubic K$_2$PtCl$_6$-type antifluorite structure with the Re ions 
forming an \textit{fcc} lattice, see Fig.\ \ref{fig:structure}a).
The structure can be viewed as equivalent to a double perovskite K$_2$$ABX_6$ in which 
the $B$ sites are occupied by Re$^{4+}$ ions while the $A$ sites correspond to 'ordered' 
vacancies.  
This material has been proposed by Streltsov and Khomskii \cite{Streltsov20} as a possible 
candidate for a spin-orbit-driven Jahn-Teller effect, arguing that it shows a series of 
structural phase transitions at lower temperature \cite{Oleary70,Armstrong80,Bertin22}. 
Recent Raman scattering results revealed the violation of cubic selection rules already 
at 300\,K \cite{Stein23}. A Curie-Weiss fit of the magnetic susceptibility \cite{Bertin22} 
yields an effective magnetic moment $\mu_{\rm eff}$\,$\approx$\,3.81\,$\mu_B$ which is 
close to the value expected for a $S$\,=\,3/2 system, 
$2\sqrt{S(S+1)}$\,$\mu_B$\,$\approx$\,3.87\,$\mu_B$.
Below the antiferromagnetic ordering temperature $T_N$\,=\,12\,K, the application of 
a large magnetic field induces weak ferromagnetism and reveals a pronounced 
magneto-elastic coupling \cite{Bertin22}. 
The structure shows nearly undistorted, unconnected ReCl$_6$ octahedra and the corresponding 
(nearly) spin-forbidden on-site $d$-$d$ excitations yield narrow features in the optical 
conductivity that allow for a most accurate determination of the excitation energies. 
In combination with the value of the cubic crystal-field splitting 10\,$Dq$ seen in RIXS, 
this allows us to obtain a reliable and accurate set of the electronic parameters. 
Our analysis is based on calculations of the local multiplet energies using 
\textsc{Quanty} \cite{Haverkort12, Haverkort16}.

The paper is organized as follows. 
Experimental aspects are described in Sect.\ \ref{sec:spectroscopy}. 
In Sect.\ \ref{sec:RIXS} and \ref{sec:assign}, we assign the features observed in RIXS 
and optical spectroscopy, respectively.\@
In optics, the spectral weight of the on-site intra-$t_{2g}$ excitations mainly stems from 
phonon-assisted processes, and the corresponding line shape and temperature-dependent spectral 
weight are discussed in Sect.\ \ref{sec:sideband} and \ref{sec:SW}, respectively. 
Overtones or double $d$-$d$ excitations are analyzed in Sect.\ \ref{sec:overtones}. 
Subtle temperature-induced effects in the optical data down to $T_N$ are presented 
in Sect.\ \ref{sec:temp}, while the temperature range below $T_N$ is addressed 
in Appendix A.\@
Results of local multiplet calculations are discussed in Sect.\ \ref{sec:theo}. 
We cover the Kanamori picture with 10\,$Dq$\,=\,$\infty$ but also derive the electronic 
parameters for finite 10\,$Dq$. We give the analytic expressions for the wavefunctions 
in the Kanamori picture, with details described in Appendix B.\@ 
This allows for a quantitative description of the effect of spin-orbit coupling.

\section{Experimental}
\label{sec:spectroscopy}

Single crystals of K$_2$ReCl$_6$ have been grown from HCl solution by controlled 
slow evaporation of the solvent. The sample batch has been thoroughly characterized 
by x-ray diffraction and measurements of the magnetic susceptibility and the specific 
heat \cite{Bertin22}. 
At 300\,K, K$_2$ReCl$_6$ exhibits the cubic K$_2$PtCl$_6$-type antifluorite structure 
with space group \textit{Fm$\bar{3}$m} and lattice parameter $a$\,=\,9.8\,\AA. 
Structural phase transitions have been observed at 111, 103, and 77\,K 
\cite{Oleary70,Armstrong80,Bertin22} and 
are accompanied by rotations and tilts of the Cl$_6$ octahedra. Note that adjacent 
Cl$_6$ octahedra are not connected in this structure, i.e., they do not share a corner, 
edge, or face. The structure turns tetragonal (\textit{P4/mnc}) at 111\,K \cite{Bertin22}, 
and monoclinic (\textit{C2/c}) at 103\,K.\ The transition from monoclinic \textit{C2/c} to 
monoclinic \textit{P2$_1$/n} is of first order, while the monoclinic angle is very close to 
90$^\circ$ in both phases \cite{Bertin22}. Finally, a magnetic phase transition to 
an antiferromagnetically ordered state occurs at $T_N$\,=\,12\,K \cite{Busey62,Smith66,Bertin22}.

To the best of our knowledge, RIXS at the Re $L_3$ edge thus far has only been reported in 
an early study on ReO$_2$ and ReO$_3$ \cite{Smolentsev11} and in the $5d^1$ and $5d^2$ 
double perovskites $A_2$$B$ReO$_6$ ($A$\,=\,Ba, Sr, Ca; $B$\,=\,Mg, Y, Cr) \cite{Paramekanti18,Yuan17,Marcaud23,Frontini23}. 
On K$_2$ReCl$_6$, resonant magnetic 
scattering has been studied at the Re $L$ edge \cite{McMorrow02}. 
Our RIXS experiments at the Re $L_3$ edge were performed at beamline ID20 
of the European Synchrotron Radiation Facility.
Incident photons from three consecutive U26 undulators were monochromatized by a Si(111) 
high-heat-load monochromator and a successive Si(311) channel-cut post-monochromator. 
Via a mirror system in Kirkpatrick-Baez geometry, the monochromatic x-ray beam was 
focused to 8\,$\times$\,50\,$\mu$m$^2$ (V\,$\times$\,H) at the sample position.
Energy loss spectra were collected with the high-energy-resolution resonant inelastic 
X-ray scattering spectrometer equipped with a 2\,m analyzer/detector arm. 
Incident $\pi$ polarization in the horizontal scattering plane was used. 
We employed the Si(9,1,1) reflection of a diced Si(11,1,1) analyzer crystal 
in conjunction with a pixelated area detector \cite{Huotari2005,Huotari2006,Moretti2013}. 
To address the resonance behavior, we have measured RIXS spectra at 300\,K with the 
incident energy in the range from 10.530 to 10.545\,keV, i.e., across the maximum of the 
Re $2p_{3/2} \rightarrow 5d$ absorption.
The overall energy resolution was 295\,meV as estimated by the full width at 
half maximum of quasielastic scattering from a piece of adhesive tape. 
The RIXS measurements were performed on a (111) surface, with (001) and (110) lying 
in the horizontal scattering plane. 
All RIXS data are corrected for the geometrical contribution to self absorption \cite{Minola15}. 
We use reciprocal lattice units for the transferred momentum $\mathbf{q}$.

Infrared and optical transmittance measurements 
were performed using a Bruker IFS 66/v Fourier-transform spectrometer. 
The energy resolution was set to 1\,cm$^{-1}$\,$\approx$\,0.12\,meV. 
We measured an as-grown sample with thickness $d$\,=\,471(5)\,$\mu$m. 
The light propagated along the cubic (111) direction.  
Using a continuous-flow $^{4}$He cryostat, the measurements were performed at 
several temperatures between 6 and 300\,K.\@

\section{Results}
\label{sect:resultsCl}

The combination of RIXS and optical spectroscopy is suited very well to examine the local electronic structure of Mott insulators such as K$_2$ReCl$_6$. RIXS at the $L_3$ edge is boosting the intensity of on-site $d$-$d$ excitations via resonant transitions between the valence $5d$ orbitals and core $2p$ states \cite{Takahashi21,Ishikawa19,Yuan17,Khan19,ReigiPlessis20,Warzanowski23,Revelli19a,Ament11}.
In contrast, optics is most sensitive to excitations that involve a change of the electric dipole moment. In Mott insulators, these are in particular inter-site excitations such as excitations across the Mott gap, here $|d^3_id^3_j\rangle\rightarrow |d^2_id^4_j\rangle$ with sites $i$ and $j$, or charge-transfer processes 
$|3p^6_{\mathrm{Cl}}5d_{\mathrm{Re}}^3\rangle\rightarrow |3p^5_{\mathrm{Cl}}5d_{\mathrm{Re}}^4\rangle$ 
\cite{Goessling08,Reul12,Vergara22}. Compared to these strong absorption bands, on-site $d$-$d$ excitations are observed as weak features \cite{Henderson,Hitchman,Rueckamp05,Benckiser08,Schmidt13,Warzanowski20,Warzanowski23}. 
In the presence of inversion symmetry, on-site $d$-$d$ excitations are parity forbidden 
by the Laporte rule. This can be circumvented by means of a phonon-assisted process 
in which the additional creation or annihilation of an odd-symmetry phonon breaks 
inversion symmetry. 
Additionally, we have to consider the spin selection rule $\Delta S$\,=\,0.
In a spin-only picture of the $t_{2g}^3$ configuration of a Re$^{4+}$ ion in K$_2$ReCl$_6$, 
the local ground state exhibits $S$\,=\,3/2 while all excited states show $S$\,=\,1/2, 
hence all excitations within the $t_{2g}^3$ manifold are spin forbidden. However, 
these excitations may acquire finite spectral weight due to spin-orbit coupling or 
in a magnetic dipole transition. 
Combining parity and spin selection rules, the spectral weight of on-site $d$-$d$ 
excitations in K$_2$ReCl$_6$ is orders of magnitude smaller than for 
strong, directly electric dipole allowed transitions. Nevertheless such weak 
on-site $d$-$d$ excitations can be studied very well in transmittance 
measurements on single crystals with an appropriate thickness, see Sect.\ \ref{sec:optics}.

\subsection{RIXS on K$_2$R\lowercase{e}C\lowercase{l}$_6$}
\label{sec:RIXS}

\begin{figure}[t]
	\centering
	\includegraphics[width=\columnwidth]{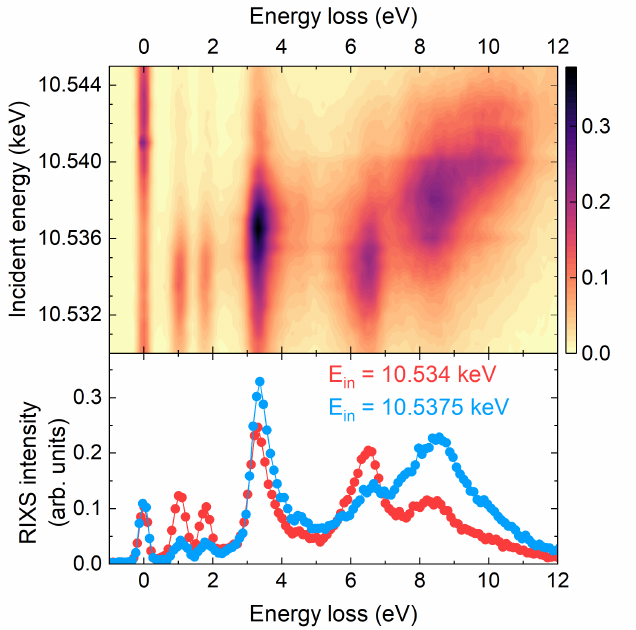}
	\caption{{\bf RIXS data of K$_2$ReCl$_6$ at $T$\,=\,300\,K.}
	Top: Resonance map of the RIXS intensity based on spectra measured with 
	different incident energy $E_{\rm in}$ for transferred momentum 
	$\mathbf{q}$\,$\approx$\,(7.4\,\,7.4\,\,5.9). 
	Bottom: RIXS spectra for $E_{\rm in}$\,=\,10.534\,keV and 10.5375\,keV, i.e., 
	at $t_{2g}$ and $e_g$ resonance, respectively.
}
\label{fig:map}
\end{figure}

The resonance behavior observed in RIXS allows us to assess the character of the 
electronic excitations in K$_2$ReCl$_6$, i.e., to distinguish intra-$t_{2g}$ features, 
$t_{2g}$-to-$e_g$ excitations, and charge-transfer excitations. The top panel of 
Fig.\ \ref{fig:map} depicts a resonance map, i.e., an intensity plot of RIXS spectra 
for different incident energies.
The data have been measured at $T$\,=\,300\,K for transferred momentum 
$\mathbf{q}$\,$\approx$\,(7.4\,\,7.4\,\,5.9). 
We find that the resonance enhancement of the RIXS intensity is peaking 
at $E_{\rm in}$\,=\,10.534\,keV and 10.5375\,keV, while the energy loss does not 
depend on $E_{\rm in}$. Cuts through the resonance map at these two incident energies 
are shown in the lower panel of Fig.\ \ref{fig:map}. 
The two resonance energies can be attributed to $t_{2g}$ resonance and $e_g$ resonance, 
i.e., enhancement of the RIXS intensity if $E_{\rm in}$ is tuned to promote a $2p$ core 
electron to either a $t_{2g}$ or an $e_g$ orbital, respectively.

The two low-energy peaks at 1.0 and 1.8\,eV display $t_{2g}$ resonance and therefore 
can be assigned to intra-$t_{2g}$ excitations. The energy resolution of 295\,meV does not 
allow us to resolve any substructure of these two peaks in the RIXS data. 
The strong feature peaking at 3.5\,eV shows $e_g$ resonance and corresponds to excitations 
from $|t_{2g}^3\rangle$ to $|t_{2g}^2e_g^1\rangle$. This assignment is supported by the 
difference between the two resonance energies, (10.5375 - 10.534)\,keV\,=\,3.5\,eV, 
which provides an approximate measure of the cubic crystal-field splitting 10\,$Dq$. 
Charge-transfer excitations	set in at about 5\,eV.\@ They correspond to promoting an electron 
from the ligand Cl $p$ shell to either Re $t_{2g}$ or $e_g$ orbitals, 
$|3p^6_{\mathrm{Cl}}5d_{\mathrm{Re}}^3\rangle\rightarrow |3p^5_{\mathrm{Cl}}5d_{\mathrm{Re}}^4\rangle$. 
Accordingly, the respective RIXS peaks at about 6.5 and 8.5\,eV exhibit $t_{2g}$ or $e_g$ resonance.

\begin{figure}[t]
	\centering
	\includegraphics[width=\columnwidth]{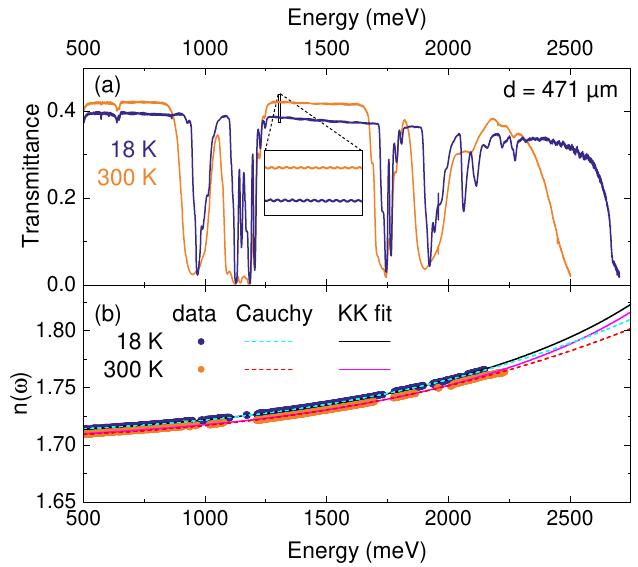}
	\caption{{\bf Optical data of K$_2$ReCl$_6$.}
	a) Transmittance $T(\omega)$ for a sample thickness $d$\,=\,471\,$\mu$m at 18 
	and 300\,K.\@  
	The data reveal weak excitations within the $t^3_{2g}$ subshell, while the onset of 
	excitations across the Mott gap at $\Delta_{\rm exp}$\,=\,2.7\,eV at 18\,K marks 
	the upper energy limit of the transparency range. 
	The inset highlights Fabry-P\'erot interference fringes.
	(b) Refractive index $n(\omega)$ extracted from Fabry-P\'erot interference fringes 
	in $T(\omega)$ (symbols). Note that $n(\omega)$ is nearly constant below the gap. 
	Dashed: Empirical fit using the Cauchy model.
	Lines: Kramers-Kronig-consistent fit using an oscillator model with an infrared-active phonon at $E_{ph}$\,=\,39\,meV and a Tauc-Lorentz oscillator with the measured gap 
	$\Delta_{\rm exp}$\,=\,2.7\,eV.
	}
	\label{fig:n}
\end{figure}

Previously, such a coexistence of $t_{2g}$ resonance and $e_g$ resonance of charge-transfer 
excitations has been reported in, e.g., $5d^2$ Ba$_2$YReO$_6$ \cite{Yuan17} and the 
sister compounds $5d^4$ K$_2$OsCl$_6$ \cite{Warzanowski23} and 
$5d^5$ K$_2$IrBr$_6$ \cite{ReigiPlessis20}. 
Compared to K$_2$OsCl$_6$ \cite{Warzanowski23}, the charge-transfer peaks are roughly 0.5 to 
0.8\,eV higher in energy in K$_2$ReCl$_6$ while the value of 10\,$Dq$ is very similar. 
Concerning the intra-$t_{2g}$ excitations, our RIXS data of K$_2$ReCl$_6$ resolve the 
two energies 1.0 and 1.8\,eV, while a splitting into four peaks at somewhat lower energies 
has been reported in RIXS on the $5d^3$ osmates Ca$_3$LiOsO$_6$ and Ba$_2$YOsO$_6$ 
\cite{Taylor17}. All five intra-$t_{2g}$ energies are revealed by our optical data, 
which we address in the next section.

\begin{figure}[t]
	\centering
	\includegraphics[width=\columnwidth]{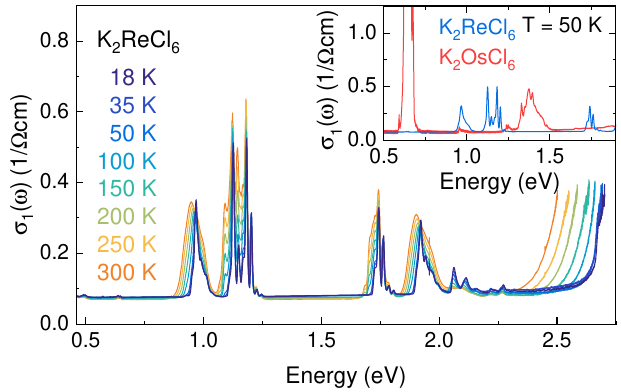}
	\caption{{\bf Optical conductivity $\sigma_1(\omega)$ of K$_2$ReCl$_6$ 
    below the Mott gap for several temperatures.} 
	The onset of excitations across the Mott gap is at 2.7\,eV at 18\,K.\@ 
	Below the gap, there are phonon-assisted intra-$t_{2g}$ excitations in the range from 
	0.9 to 2\,eV.\@ Above 2\,eV, we observe weak overtones, i.e., inter-site pair excitations 
	that correspond to two intra-$t_{2g}$ excitations on neighboring sites. 
	Inset: Comparison of intra-$t_{2g}$ excitations of K$_2$ReCl$_6$ and its $d^4$ sister 
	compound K$_2$OsCl$_6$ \cite{Warzanowski23}. From the latter, a small constant offset 
	has been subtracted.
}
\label{fig:sigma_all}
\end{figure}

\subsection{Optical conductivity of K$_2$ReCl$_6$}
\label{sec:optics}

The experimental task is to obtain the complex optical conductivity 
$\sigma(\omega)$\,=\,$\sigma_1(\omega) + i \sigma_2(\omega)$ or, equivalently, 
the complex index of refraction $N(\omega)$\,=\, $n(\omega) + i\kappa(\omega)$, 
with $\sigma_1 \propto n k$. 
To address the very weak on-site $d$-$d$ absorption features in the frequency range 
below the Mott gap, we employ the sensitivity of the transmittance $T(\omega)$ which 
depends exponentially on $\kappa(\omega)$, see Fig.\ \ref{fig:n}.
In contrast to $\kappa(\omega)$, the real part $n(\omega)$ is not very sensitive to weak 
absorption bands. In the transparency range $\kappa \ll n$, Fabry-P\'erot interference 
fringes are observed in $T(\omega)$ due to multiple reflections within the sample. 
From the fringes we determine the optical path length $n(\omega) d$ and hence 
the real part $n(\omega)$, see Fig.\ \ref{fig:n}.
As expected, $n(\omega)$ shows the nearly constant behavior that is typical for an 
insulator below the gap. Sufficiently far above the phonon range, the small positive dispersion 
can be described empirically by the Cauchy model, $n$\,=\,$\alpha + \beta/\lambda^2 + \gamma/\lambda^4$ where $\lambda$ denotes the wavelength and $\alpha$, $\beta$, and 
$\gamma$ are fit parameters (dashed lines in Fig.\ \ref{fig:n}).

\subsubsection{Assignment of intra-$t_{2g}$ excitations}
\label{sec:assign}

The real part of the optical conductivity $\sigma_1(\omega)$ of K$_2$ReCl$_6$ 
is plotted in Fig.\ \ref{fig:sigma_all}. We focus on temperatures larger than 
$T_N$\,=\,12\,K  to avoid the additional complexity arising from magnetic order, 
see Appendix A.\@ We observe three different kinds of excitations:
the onset of excitations across the Mott gap at about 2.7\,eV at 18\,K, a series of 
weak phonon-assisted intra-$t_{2g}$ excitation bands between 0.9\,eV and 2\,eV 
that are in agreement with the corresponding RIXS peaks centered at 1.0 and 1.8\,eV,
and even weaker features above 2\,eV that can be attributed to double excitations, i.e., 
combinations and overtones of the intra-$t_{2g}$ excitations, as discussed below. 
Overall, the values of $\sigma_1(\omega)$ are very small below the gap, reflecting the 
infrared-forbidden character of the intra-$t_{2g}$ excitations \cite{Henderson,Hitchman}. 
The weak spectral weight of excitations that are both spin-forbidden and parity-forbidden 
is most strikingly illustrated by data on compounds where both spin-forbidden and spin-allowed 
excitations are observed, such as $3d^2$ VOCl \cite{Benckiser08} or 
$3d^3$ Cr compounds \cite{Schmidt13}. Furthermore, it is instructive to compare K$_2$ReCl$_6$ with 
the $5d^4$ $J$\,=\,0 compound K$_2$OsCl$_6$ \cite{Warzanowski23}, see inset of 
Fig.\ \ref{fig:sigma_all}. In the absence of spin-orbit coupling, the cubic $t_{2g}^4$ 
configuration would show an $S$\,=\,1 ground state and $S$\,=\,0 excited states, 
i.e., spin-forbidden excitations. Most of the features in $J$\,=\,0 K$_2$OsCl$_6$ are as weak 
as in K$_2$ReCl$_6$. 
However, K$_2$OsCl$_6$ also exhibits a stronger band around 0.6\,eV that corresponds to 
the excitation from $J$\,=\,0 to 2, reflecting the prominent role of strong spin-orbit coupling 
for this absorption feature. This comparison gives a first hint for a smaller effect of 
spin-orbit coupling for the $t_{2g}^3$ configuration of K$_2$ReCl$_6$.

\begin{figure}[t]
	\centering
	\includegraphics[width=\columnwidth]{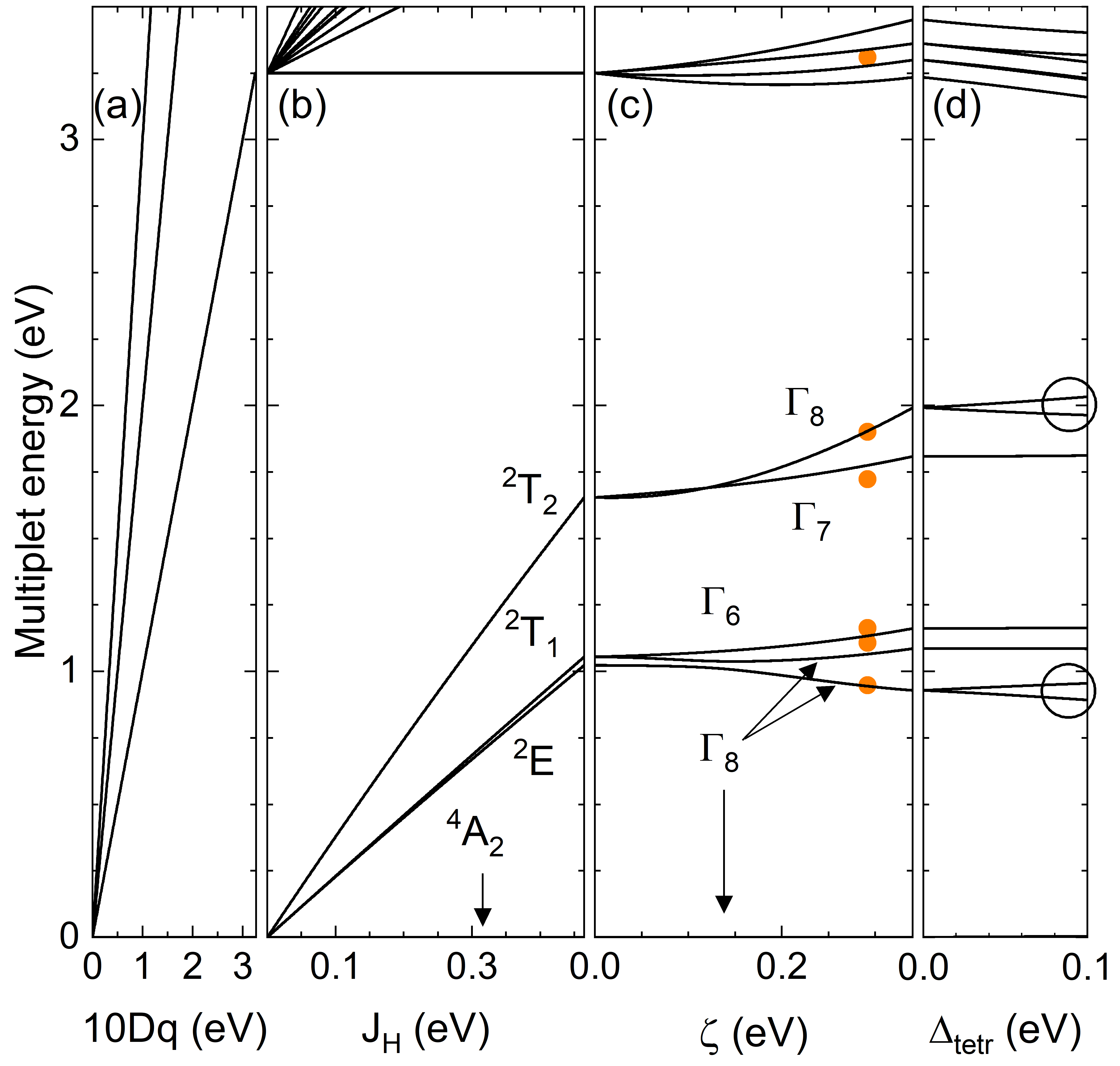}
	\caption{{\bf Energy level diagram for the $d^3$ configuration.} 
	a)-c) Using \textsc{Quanty} \cite{Haverkort12, Haverkort16}, the cubic multiplet energies 
	were calculated 
	as a function of cubic crstal-field splitting 10\,$Dq$, spin-orbit coupling $\zeta$, and 
	interelectronic Coulomb interaction in terms of the Slater integrals $F^2$ and $F^4$. 
	The latter is captured by $J_{\rm H}$\,=\,$1/14 (F^2+F^4)$. 
	Panel a) uses $J_{\rm H}$\,=\,$\zeta$\,=\,0 and shows the $t_{2g}^3$ states at $E$\,=\,0 and $t_{2g}^{3-n}e_g^n$ states at higher energy. 
	b) Coulomb interactions lift the degeneracy of the $t_{2g}^3$ states.
	c) Effect of $\zeta$. Orange symbols: Intra-$t_{2g}$ energies taken from $\sigma_1(\omega)$ and the $t_{2g}$-to-$e_g$ excitation energy of 3.3\,eV observed in RIXS.\@ Their position on the $\zeta$ axis marks the value obtained from a fit (see main text). 
	For the $e_g$ states, the maximum of the broadened RIXS response was chosen, lying between the second and third energy levels. 
	d) Effect of a tetragonal crystal field. Circles highlight the splitting of the lowest and 
	highest intra-$t_{2g}$ excitations.
	}
	\label{fig:energy}
\end{figure}

Our focus is on the intra-$t_{2g}$ excitations of K$_2$ReCl$_6$. In total, there are 20 
$t_{2g}^3$ states. In cubic symmetry and neglecting spin-orbit coupling, these are split 
by Coulomb interactions into four multiplets, i.e., the high-spin $S$\,=\,3/2 $^4A_2$ 
ground state and the three low-spin $S$\,=\,1/2 excited states $^2E$, $^2T_1$, and $^2T_2$, 
see Fig.\ \ref{fig:energy}.
In the $t_{2g}$-only Kanamori scheme, i.e., for a cubic crystal-field splitting 
10\,$Dq$\,=\,$\infty$, the excitation energies are $3J_{\rm H}^{\rm eff}$ for the $^2E$ and 
$^2T_1$ multiplets and $5J_{\rm H}^{\rm eff}$ for $^2T_2$ \cite{Hoggard81,Georges13,Zhang17}, 
where $J_{\rm H}^{\rm eff}$ denotes Hund's coupling in the Kanamori scheme \cite{Georges13} 
(see Sect.\ \ref{sec:Kanamori}).
From the peak energies observed in RIXS, 1.0 and 1.8\,eV, we obtain a rough estimate 
$J_{\rm H}^{\rm eff}$\,$\approx$\,0.35\,eV.\@ 
For finite 10\,$Dq$, in a model considering the entire $d$ shell \cite{Georges13}, 
this corresponds to $J_{\rm H}$\,$\approx$\, 
$J_{\rm H}^{\rm eff}/0.77$\,$\approx$\,0.45\,eV, 
see Fig.\ \ref{fig:energy}b).
\\

\begin{figure}[t]
	\centering
	\includegraphics[width=\columnwidth]{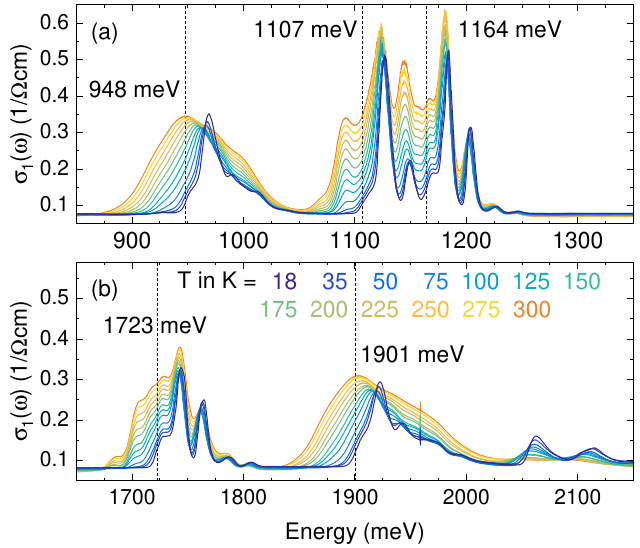}
	\caption{{\bf Temperature dependence of intra-$t_{2g}$ excitations.} 
	Both panels depict $\sigma_1(\omega)$ over a range of 0.5\,eV.\@ The multi-peak 
	structure and the temperature-driven increase of spectral weight indicate the 
	phonon-assisted character, see Fig.\ \ref{fig:shift}. 
	Vertical dashed lines denote the bare electronic energies $E_{0,i}$  
	for the excited states 
	$\Gamma_8$ at 948\,meV, 
	$\Gamma_8$ at 1107\,meV,  
	$\Gamma_6$ at 1164\,meV, 
	$\Gamma_7$ at 1723\,meV, and 
	$\Gamma_8$ at 1901\,meV.\@
	The tiny feature at 1959\,meV (633\,nm) is an artifact from a HeNe laser used to 
	calibrate the energy of the Fourier spectrometer.
}
\label{fig:sigma_t2g}
\end{figure}

Finite 10\,$Dq$ lifts the degeneracy between $^2E$ and $^2T_1$, while spin-orbit 
coupling splits both $^2T_1$ and $^2T_2$ into a quartet and a doublet. Altogether, 
this yields five excited states within the $t_{2g}^3$ manifold. In Bethe notation, 
the ground state is given by $\Gamma_8$ while the five excited states are described 
by $\Gamma_8$, $\Gamma_8$, $\Gamma_6$, $\Gamma_7$, and $\Gamma_8$, 
see Fig.\ \ref{fig:energy} and Appendix B.\@
In the optical data, the spin-forbidden character of the intra-$t_{2g}$ excitations 
yields narrow absorption lines that allow for an accurate determination of all five 
electronic excitation energies $E_{0,i}$ for $i$\,=\,1-5 \cite{Pross74,Yoo87}. 
As explained in the next paragraph, we find $E_{0,i}$\,=\,948, 1107, 1164, 1723, 
and 1901\,meV, see Fig.\ \ref{fig:sigma_t2g}.

\begin{figure}[t]
	\centering
	\includegraphics[width=\columnwidth]{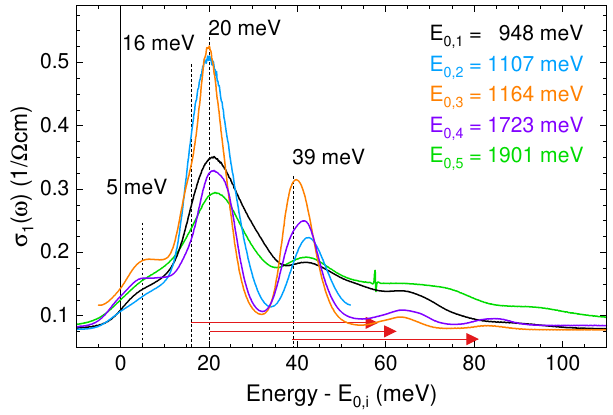}
	\caption{{\bf Common peak structure of intra-$t_{2g}$ excitations at 18\,K.} 
	Data of the five absorption bands $i$\,=\,1-5, see Fig.\ \ref{fig:sigma_t2g}, are 
	plotted as a function of $E\! - \! E_{0,i}$ to highlight the common energies of the 
	symmetry-breaking phonons with $E_{\rm ph}$\,=\,5, 16, 20, and 39\,meV (vertical 
	dotted lines) that yield peaks at $E_{0,i} \! + \!  E_{\rm ph}$. Moreover, the plot 
	shows the Franck-Condon phonon sidebands at 
	$E_{0,i} \! + \! E_{\rm ph} \! + \! E_{a_{1g}}$ with $E_{a_{1g}}$\,$\approx$\,44\,meV (arrows). Note that the excitations at 
	$E_{0,2}$\,= 1107\,meV and $E_{0,3}$\,=\,1164\,meV  
    are close in energy (light blue and orange). To disentangle the corresponding 
    features, the lines have been cut at 1159\,meV, i.e., at 52\,meV for $E_{0,2}$ 
    and -5\,meV for $E_{0,3}$. 
	The corresponding plot for 6\,K, below $T_N$, is given in Appendix A.
}
\label{fig:shift}
\end{figure}

\subsubsection{Sideband features of the intra-$t_{2g}$ absorption bands}
\label{sec:sideband}

For each absorption band, $\sigma_1(\omega)$ shows a series of peaks, 
see Fig.\ \ref{fig:sigma_t2g}. However, the energies $E_{0,i}$ can be identified 
based on the common peak structure that becomes evident by plotting the five 
intra-$t_{2g}$ absorption bands on the shifted energy axis $E - E_{0,i}$, 
as done in Fig.\ \ref{fig:shift} for 18\,K and in Appendix A for 6\,K, below $T_N$.
These plots illustrate the existence of the following three distinct mechanisms.
Firstly, we observe very weak magnetic dipole transitions at $E_{0,i}$ \cite{Pross74,Yoo87}, 
where $E_{0,i}$ denotes the zero-phonon electronic energy of band $i$. 
Secondly, the spectra show phonon-assisted excitations 
\cite{Ballhausen,Henderson,Hitchman,Rueckamp05}
at $E_{0,i}\! + \! E_{\rm ph}$ with phonon energies $E_{\rm ph}$ of about 
5, 16, 20, and 39\,meV.\@ These phonon energies agree with the temperature dependence 
of the spectral weight, as discussed below. Thirdly, vibronic Franck-Condon-type 
absorption peaks with smaller spectral weight are observed at 
$E_{0,i}+E_{\rm ph} \! + \! E_{a_{1g}}$ with $E_{a_{1g}}$$\approx$\,44\,meV.

The upper three energies $E_{\rm ph}$\,=\,16, 20, and 39\,meV 
of symmetry-breaking phonon modes can be motivated by considering 
a single regular ReCl$_6$ octahedron. 
It exhibits three odd-symmetry normal modes of vibration that break the inversion symmetry 
on the Re site and hence contribute to phonon-assisted absorption \cite{Pross74,Yoo87}. 
The additional mode at about 5\,meV has to be identified as a lattice phonon mode. 
Far-infrared data reported a phonon at 5.3\,meV that becomes infrared active by backfolding 
below 103\,K \cite{Oleary70}. Note that the symmetry-breaking modes do not have to be at the 
$\Gamma$ point, i.e., they do not have to be infrared active.

\begin{figure}[t]
	\centering
	\includegraphics[width=\columnwidth]{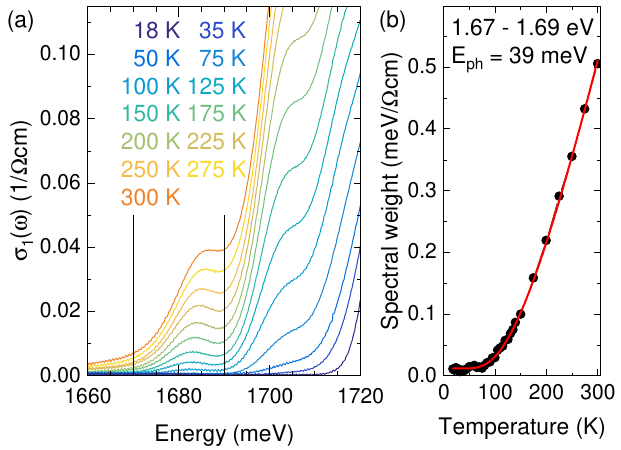}
	\caption{{\bf Temperature dependence of $\sigma_1(\omega)$ around 1.7\,eV.\@} 
	(a) Close-up of $\sigma_1(\omega)$ highlighting the thermal population of the phonon 
	mode with $E_{\rm ph}$\,=\,39\,meV that yields an intra-$t_{2g}$ excitation at 
	$E_{0,4}-E_{\rm ph}$ for $E_{0,4}$\,=\,1723\,meV.\@ 
	For each temperature, an offset has been subtracted that was fixed outside the 
	absorption band, as described for Fig.\ \ref{fig:weight}. 
	Vertical black lines denote the energy range used for the integration to calculate 
	the spectral weight. 
	(b) Spectral weight of the phonon-assisted feature shown in a) (symbols). 
	Red line: Fit using a Bose occupation factor with $E_{\rm ph}$\,=\,39\,meV.
	}
	\label{fig:39mev}
\end{figure}

The phonon-assisted scenario is supported by the temperature dependence of the 
spectral weight. At low temperature such as 18\,K, the spectra show absorption peaks 
at $E_{0,i}\! + \! E_{\rm ph}$ in which a phonon and an electronic excitation are 
excited simultaneously. 
With increasing temperature, the spectral weight grows at $E_{0,i}\! - \! E_{\rm ph}$  
due to phonon-annihilation contributions, giving rise to a temperature-induced increase 
of the spectral weight in particular on the low-energy side of each band. 
For, e.g., $E_{0,4}$\,= 1723\,meV, the integrated spectral weight around 
$E_{0,4}$-39\,meV reveals a Bose factor that agrees with a phonon energy of 39\,meV, 
see Fig.\ \ref{fig:39mev}. 
A quantitative analysis of the total integrated spectral weight of the absorption bands 
will be discussed in Sect.\ \ref{sec:SW}.

In contrast to the peaks at $E_{0,i} \pm  E_{\rm ph}$ that reflect the phonon-assisted 
excitation mechanism, the vibronic Franck-Condon type phonon sidebands at 
$E_{0,i}+E_{\rm ph}+E_{a_{1g}}$ arise due to 
the finite coupling of the electronic excitations to the lattice.  
A change of the orbital occupation upon electronic excitation implies that the lattice 
is not necessarily in its ground state anymore. This yields a vibronic character of the 
modes, i.e., mixed electronic and vibrational character, which drives phonon sidebands according 
to the Franck-Condon principle \cite{Ballhausen,Rueckamp05}. 
For a $t_{2g}^3$ configuration in a regular octahedron, both the electronic ground state and 
the excited states exhibit cubic symmetry. Preserving cubic symmetry, the lattice may relax via 
a Raman-active breathing mode of the ReCl$_6$ octahedron with $a_{1g}$ symmetry, changing 
the Re-Cl distance. In Raman scattering, the $a_{1g}$ mode has been observed at 
42\,meV \cite{Stein23}, 
in good agreement with our result. Note that our optical data probe the Raman mode in an 
electronically excited state, which may cause small shifts of the phonon energy.  
Moreover, the local character of the excitation averages over the dispersion 
across the Brillouin zone.

\subsubsection{Spectral weight}
\label{sec:SW}

\begin{figure}[t]
	\centering
	\includegraphics[width=\columnwidth]{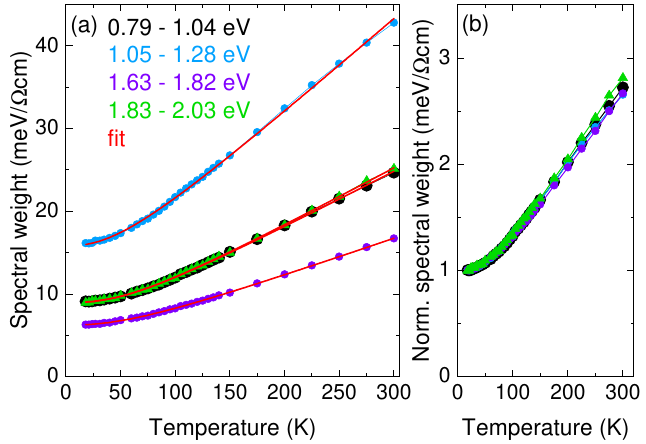}
	\caption{{\bf Spectral weight as function of temperature.} 
	(a) Spectral weight (symbols) of the absorption bands with $E_{0,i}$ within the given 
	integration ranges. Note that the data of the close lying bands with $E_{0,2}$ and 
	$E_{0,3}$ have been integrated together (blue). The data for 
	$E_{0,1}$\,=\,948\,meV (black) and $E_{0,5}$\,=\,1901\,meV (green) nearly fall 
	on top of each other. In each case, the data show the characteristic temperature 
	dependence of a phonon-assisted process, as shown by the fits (red lines) 
	following Eq.\ (\ref{eq:weight}) with four 
	phonon modes at $E_{{\rm ph},k}$\,=\,5, 16, 20, and 39\,meV.\@ 
	For each temperature, a linear offset has been subtracted from $\sigma_1(\omega)$ 
	that was fixed outside the absorption bands, i.e., at 755 and 1275\,meV for $E_{0,1}$ to 
	$E_{0,3}$ and at 1400 and 2040\,meV for the upper two bands. 
	(b) The normalized spectral weights of the four integration ranges plotted in a) collapse 
	to a single curve, corroborating the common phonon energies.
	}
	\label{fig:weight}
\end{figure}

Figure \ref{fig:weight}a) shows the integrated spectral weight of the different 
absorption bands as a function of temperature. The plot depicts data for four 
different frequency ranges. Three of them correspond to the distinct absorption 
bands around $E_{0,1}$, $E_{0,4}$, and $E_{0,5}$, while the fourth data set shows 
the cumulative spectral weight of the two close-lying bands around 
$E_{0,2}$\,=\,1107\,meV and $E_{0,3}$\,=\,1164\,meV, 
see Fig.\ \ref{fig:sigma_t2g}.
The integration ranges are given in Fig.\ \ref{fig:weight}.    
They were chosen sufficiently large to capture the spectral weight of both phonon-creating and 
phonon-annihilating processes. 
For a phonon-assisted process, we can describe the spectral weight of a feature located between $\omega_1$ and $\omega_2$ as   
\begin{align}
SW=\int_{\omega_1}^{\omega_2}\sigma_1(\omega)\mathrm{d}\omega = \alpha+\sum_k \beta_k \coth\left(\frac{E_{\mathrm{ph},k}}{2k_{\mathrm{B}}T}\right),
\label{eq:weight}
\end{align}
where $E_{\mathrm{ph},k}$ denotes the energies of the odd, symmetry-breaking phonon modes 
and $\alpha$ and $\beta_k$ are fit parameters. 
Based on the four phonon energies $E_{\mathrm{ph},k}$\,=\,5, 16, 20, and 39\,meV 
discussed in connection with Fig.\ \ref{fig:shift}, we find good agreement between model 
and data, see Fig.\ \ref{fig:weight}.
Furthermore, the curves of the normalized spectral weight nearly lie on top of each other, 
with the largest deviation of 5\,\% observed at 300\,K, see Fig.\ \ref{fig:weight}b).
This stresses the common energies $E_{\mathrm{ph},k}$ of the relevant phonon modes.

\begin{figure}[t]
	\centering
	\includegraphics[width=\columnwidth]{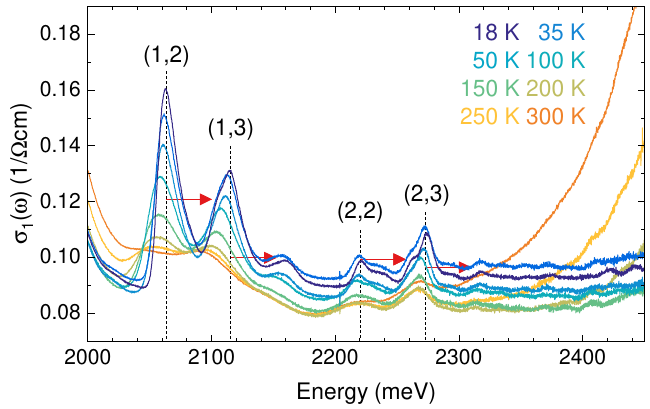}
	\caption{{\bf Intersite $d$-$d$ overtones in $\sigma_1(\omega)$ of K$_2$ReCl$_6$.} 
	Above the highest intra-$t_{2g}$ excitation with $E_{0,5}$\,=\,1901\,meV, 
	we observe peaks at energies that (nearly) coincide with the sum $E_{0,i}+E_{0,j}$ 
	of two intra-$t_{2g}$ excitation energies. 
	Hence these features are not phonon assisted but directly infrared active, in agreement 
	with their temperature dependence.  
	Still also these higher-order excitations show a Franck-Condon sideband of the $a_{1g}$ 
	breathing mode with $E_{a_{1g}}$\,=\,44\,meV (arrows, cf.\ Fig.\ \ref{fig:shift}). 
}
\label{fig:double}
\end{figure}

\subsubsection{Overtones or double $d$-$d$ excitations}
\label{sec:overtones}

According to RIXS, excitations to $e_g$ orbitals require an energy larger than 3\,eV, as 
discussed above. Having identified the five expected intra-$t_{2g}$ excitations below 
2\,eV, the series of weak absorption features between 2.0 and 2.4\,eV at first sight 
comes as a surprise, see Fig.\ \ref{fig:double}. 
The key to their assignment as double $d$-$d$ excitations, with the two excitations 
located on adjacent sites, is provided by the peak energies. At 18\,K, the values of 
2063, 2115, 2221, and 2273\,meV agree within 2 to 8\,meV with the sums 
$E_{0,i} \!  + \! E_{0,j}$ for $(i,j)$\,=\,(1,2), (1,3), (2,2), and (2,3),  
respectively, revealing the combination and overtone character \cite{Bettinelli88}.
Note that the lowest peak of this series is expected at 
$2E_{0,1}$\,=\,1896\,meV, overlapping with the band of the $\Gamma_{8}$ term 
around $E_{0,5}$\,=\,1901\,meV.\@ 
Like the single-site $d$-$d$ excitations at $E_{0,i}$, these overtones show a 
Franck-Condon-type phonon sideband that corresponds to the 44\,meV $a_{1g}$ mode, 
see arrows in Figs.\ \ref{fig:double} and \ref{fig:shift}.

The overtone peak energies may deviate from the sum of two single-site excitation 
energies due to interaction effects. In K$_2$ReCl$_6$, these deviations do not exceed 
8\,meV or 0.4\,\%, a remarkably small value that reflects once more the small coupling 
of these intra-$t_{2g}$ excitations to the lattice. The excitation energies are mainly 
determined by $J_{\rm H}$ and $\zeta$, and even on two adjacent, unconnected ReCl$_6$ 
octahedra the two excitations hardly interact. Similar overtones of intra-$t_{2g}$ 
excitations have been reported in the $5d$ sister compound 
K$_2$OsCl$_6$ \cite{Warzanowski23}, in $3d$ orbitally ordered 
YVO$_3$ \cite{Benckiser08}, and in the $4d$ Kitaev material \mbox{$\alpha$-RuCl$_3$}.  
In the latter, both double and triple excitations were observed \cite{Warzanowski20}.

With the peak energies coinciding with the sums $E_{0,i}\!+\!E_{0,j}$ of the purely electronic  
zero-phonon energies, these features are directly infrared allowed, i.e., not phonon assisted. 
This is supported by their temperature dependence, which lacks the temperature-driven increase 
of spectral weight described by Eq.\ (\ref{eq:weight}).
Previously, the finite spectral weight of these overtones has been attributed to 
quadrupole-quadrupole interactions between neighboring Re sites \cite{Bettinelli88}. 
In this scenario, the spectral weight is enhanced by some mixing between the overtones 
and the single-site excitation at $E_{0,5}$\,=\,1901\,meV, and this mixing has been 
claimed to explain the decrease of intensity of the overtones with increasing separation from $E_{0,5}$. 
In the sister compound K$_2$OsCl$_6$, however, the intensity of the overtones is not correlated 
with the distance in energy to single-site $d$-$d$ excitations \cite{Warzanowski23}.
Alternatively, we consider a hopping-based mechanism related to superexchange. 
Lorenzana and Sawatzky \cite{Lorenzana95a, Lorenzana95b} described an analogous process 
for magnetic excitations. 
In an antiferromagnet, the exchange of two spins on adjacent sites is equivalent to a double 
spin flip. This excitation may generate an electric dipole moment if inversion symmetry 
is broken on the bond between the two sites. In other words, a finite dipole moment arises 
if the matrix elements for the corresponding electron transfer from site $i$ to site $j$ 
and \textit{vice versa} are different. In the case of inversion symmetry, finite 
spectral weight of such a double spin flip can be caused by a phonon-assisted process, 
the so-called bi-magnon-plus-phonon absorption \cite{Lorenzana95a, Lorenzana95b,Windt01}. 
For double orbital excitations, both the direct process and the phonon-assisted version 
have been observed \cite{Benckiser08,Warzanowski20}. 
In the cubic phase above 111\,K, K$_2$ReCl$_6$ shows inversion symmetry on the midpoint 
between two adjacent Re sites \cite{Bertin22}. We attribute the larger spectral weight 
at low temperature to the breaking of inversion symmetry in the low-temperature phases. 
The finite spectral weight above 111\,K tentatively can be attributed to fluctuations of 
octahedral rotations that according to Raman scattering \cite{Stein23} give rise to a 
low-energy continuum of excitations that extends up to about 12\,meV, both at low temperature 
and at 300\,K.\@

\subsubsection{Differences in line shape and temperature dependence}
\label{sec:temp}

We established that the five absorption bands around $E_{0,i}$ share a common temperature dependence of the spectral weight, see Fig.\ \ref{fig:weight}, and a common peak structure with $\sigma_1(\omega)$ at low temperature peaking at $E_{0,i}+E_{\rm ph}$ and 
$E_{0,i}+E_{\rm ph}+E_{a_{1g}}$ with common phonon energies, see Fig.\ \ref{fig:shift}. 
However, a closer look at the line shape reveals two different types of behavior that distinguishes the absorption bands $i$\,=\,1 and 5 from those with $i$\,=\,2-4. 
The differences are apparent in the width of the individual peaks, 
the spectral weight of the Franck-Condon side bands at $E_{0,i}+E_{\rm ph}+E_{a_{1g}}$, 
and the precise temperature dependence of the peak energies and peak heights. 
We address these points in the following.  

\begin{figure}[t]
	\centering
	\includegraphics[width=\columnwidth]{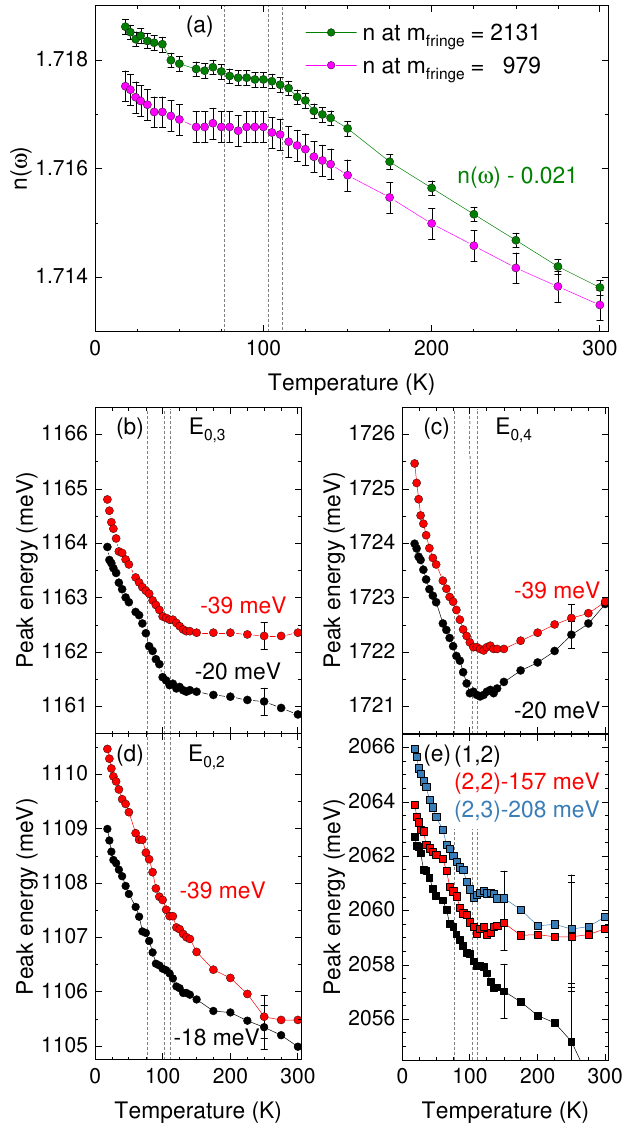}
	\caption{{\bf Temperature dependence of the refractive index $n(\omega)$ and of 
			the intra-$t_{2g}$ energies.} 
		(a) $n(\omega)$ extracted from the energies of two Fabry-P\'erot interference fringes at around 0.75 and 1.6\,eV with order $m_{\mathrm{fringe}}$\,=\,979 and 2131, respectively. 
		The curve of $n$ of the latter has been shifted down by 0.021 to facilitate comparison. 
		(b)-(d) Peak energies $E_{0,i}+E_{\rm ph}$ for $i$\,=\,2, 3, and 4 and two different phonon modes. Symbols have been shifted as indicated in the panels.
		(e) Peak energies of the overtones at $E_{0,i}+E_{0,j}$ for $(i,j)$\,=\,(1,2), (2,2), and (2,3).
		Vertical dashed lines indicate the temperatures of the structural phase transitions.
	}
	\label{fig:temp}
\end{figure}

The width of the individual peaks is larger for $i$\,=\,1 and 5 than for $i$\,=\,2-4. Below 111\,K we have to consider deviations from cubic symmetry. Figure \ref{fig:energy}d) shows that, e.g., a tetragonal crystal field yields clear splittings for $i$\,=\,1 and 5. In contrast, deviations from cubic symmetry cannot split the Kramers doublets for $i$\,=\,3 or 4. 
For the quartet $i$\,=\,2 we calculate a tiny splitting for finite $\zeta$. This band is related to the $^2E$ multiplet for $\zeta$\,=\,0, which does not split in a tetragonal crystal field. 
The non-cubic symmetry hence is a plausible explanation for the larger width for $i$\,=\,1 and 5. 
Note that our optical data do not resolve a peak splitting, we only find a larger width. 
Therefore, the non-cubic crystal-field splitting has to be small. 
Similarly, Raman data of the phonon modes at low temperature could not resolve a peak splitting \cite{Stein23}. Instead, the Raman data show a low-energy continuum that points towards rotary fluctuations of the orientation of the ReCl$_6$ octahedra. This scenario agrees with an enhanced peak width in $\sigma_1(\omega)$ for $i$\,=\, 1 and 5.

The spectral weight of the Franck-Condon-type phonon sidebands at $E_{0,i}+E_{\rm ph}+E_{a_{1g}}$ is larger for $i$\,=\,1 and 5, see Fig.\ \ref{fig:shift}. This also affects the line shape at elevated temperatures. For $i$\,=\,3 and 4, the spectral weight with increasing temperature mainly rises on the low-energy side (see above). We expect the same behavior for $i$\,=\,2 but this is covered by the overlap with $i$\,=\,3. 
In contrast, the spectral weight for $i$\,=\,1 and 5 is also enhanced on the high-energy side, 
in the range of the Franck-Condon sidebands.  
These even-symmetry Franck-Condon-type phonon sidebands should not be confused with 
the odd-symmetry mode that breaks the inversion symmetry in the phonon-assisted process. 
If an electronic excitation yields a sizable change of both the orbital occupation and 
the corresponding charge distribution, the phonon sidebands dominate the spectral weight 
and determine the line shape in the optical data \cite{Rueckamp05}. 
Moreover, the existence of several phonon modes, all of them showing dispersion, typically 
washes out the detailed sideband structure in crystalline samples, giving rise to a 
broad featureless peak in $\sigma_1(\omega)$ \cite{Rueckamp05,Benckiser08}.  
The broader line shape of $i$\,=\,1 and 5 in combination with the larger spectral weight of the Franck-Condon sidebands thus indicates that these excitations correspond to a stronger change of the orbital occupation. 
In contrast, a nearly pure spin flip with little change of the orbital occupation exhibits a very small coupling to the lattice and the spectral weight in the Franck-Condon sidebands is suppressed. In $5d$ K$_2$ReCl$_6$ with well-separated ReCl$_6$ octahedra, the very narrow peaks for $i$\,=\,2-4 underline the marginal coupling to the lattice. 
The data for $i$\,=\,2-4 qualify as a textbook example of on-site $d$-$d$ excitations with a predominant spin-flip character.

Concerning the temperature dependence, Streltsov and Khomskii \cite{Streltsov20} speculated 
that one of the phase transitions of K$_2$ReCl$_6$ could be related to a spin-orbit-induced 
Jahn-Teller splitting. In the optical data, temperature-related effects are rather subtle
above $T_N$. For instance Fig.\ \ref{fig:sigma_t2g} does not show pronounced qualitative changes of 
$\sigma_1(\omega)$ throughout the entire temperature range, and the spectral weight shows a smooth evolution with temperature, see Fig.~\ref{fig:weight}a). 
The structural changes hence hardly affect the matrix elements of the phonon-assisted processes. 
The refractive index $n(\omega)$ in the studied frequency range is governed by the spectral weight 
of the higher-lying electronic interband excitations. 
The temperature dependence of $n(\omega)$ is exemplified for two frequencies in 
Fig.\ \ref{fig:temp}a). The data suggest small changes of the slope 
$\partial n/\partial T$ but do not show strong effects.

The clearest signatures of deviations from cubic symmetry below 111\,K are detected 
in the peak energies. We focus on the bands with $i$\,=\,2 to 4 since these allow to track the individual peaks up to 300\,K.\@
In Fig.\ \ref{fig:temp}b)-d) we compare the peak energies $E_{0,i}+E_{\rm ph}$, 
considering two different phonon energies $E_{\rm ph}$. 
In all three panels, the energies continuously harden by 2.5 to 3\,meV from the highest 
phase transition temperature 111\,K down to low temperature.\@ 
Note that the shift amounts to about 0.2\,\% of the energy, a tiny effect. 
In each panel, both curves show a very similar shift, even though the phonon energies differ 
by about a factor of two. Moreover, the phonon energies in general are expected to exhibit 
a larger softening above 100\,K due to thermal expansion. We thus conclude that the observed 
behavior below 111\,K predominantly reflects the temperature dependence of the electronic 
energies $E_{0,i}$. 
This is corroborated by the peak energies of the double excitations at $E_{0,1}+E_{0,2}$, 
$2 E_{0,2}$, and $E_{0,2}+E_{0,3}$, see Fig.\ \ref{fig:temp}e). 
The temperature-induced shift of the overtones below 111\,K amounts to about 5\,meV, i.e., 
it is twice as large as the shift of the single modes. 
This common increase of the excitation energies $E_{0,i}$ strongly suggests a lowering of 
the ground state energy by about 3\,meV from 111\,K to low temperature.\@

\begin{figure}[t]
	\centering
	\includegraphics[width=0.47\columnwidth]{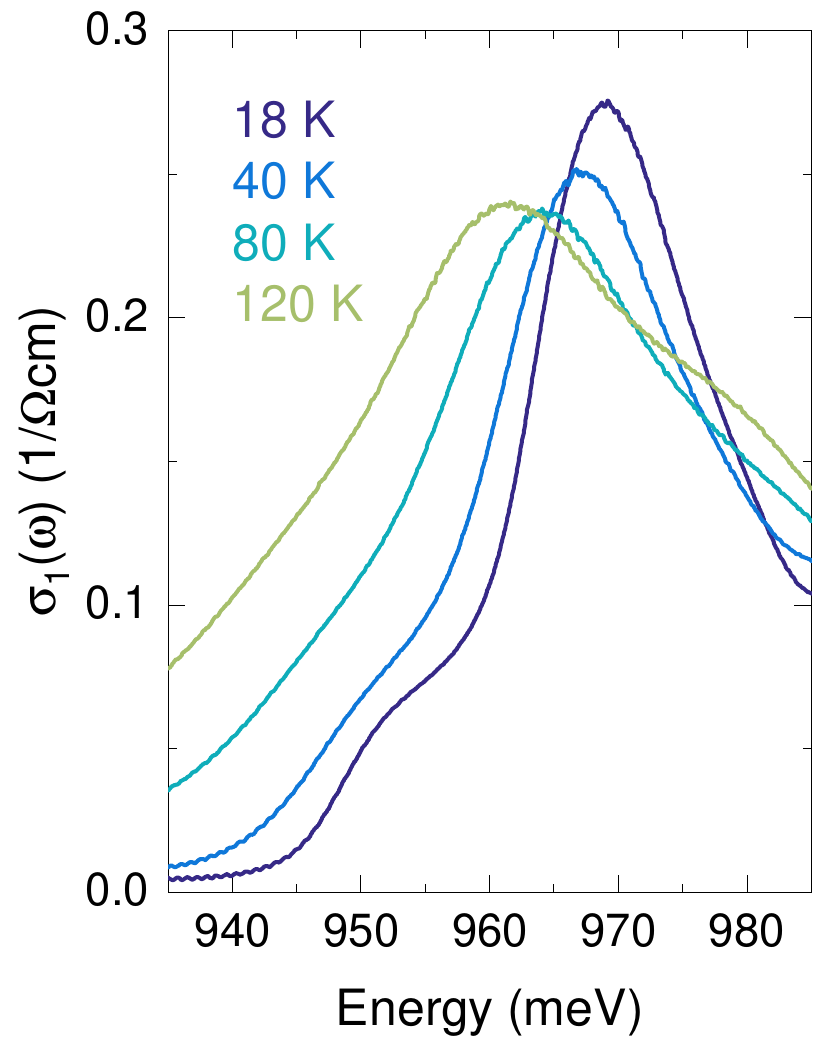}
	\hfill
	\includegraphics[width=0.49\columnwidth]{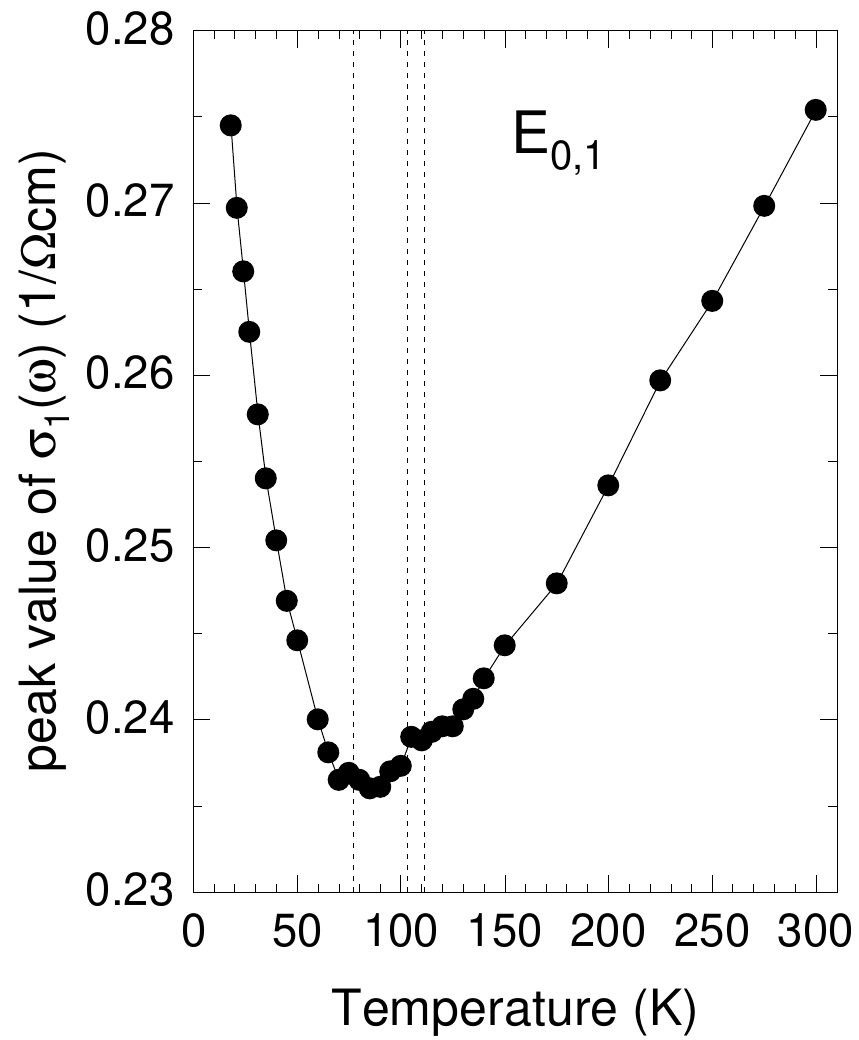}
	\caption{{\bf Value of $\sigma_1(\omega)$ at the peak maximum for the band around $E_{0,1}$\,=\,948\,meV.\@} 
	Left: Zoom in on $\sigma_1(\omega)$ around the peak at 
	$E_{0,1}+E_{\rm ph}$. A background has been subtracted, as discussed for Fig.\ \ref{fig:weight}.  
	Right: The increase of the peak maximum at high temperature reflects the increase of the spectral weight, see Fig.\ \ref{fig:weight}. In view of the spectral weight, the opposite behavior of the peak maximum 
	below 77\,K indicates a change of the linewidth. 
	}
	\label{fig:height}
\end{figure}

The peak energies of the bands with $i$\,=\,1 and 5 show larger shifts as function of temperature. For these bands, however, the larger coupling to the lattice implies a larger role of phonons for the line shape, impeding the determination of the temperature dependence of the pure electronic energies. Instead, we address the peak height, i.e., the value of $\sigma_1(\omega)$ at the peak. For $i$\,=\,2-4, the peak height increases with increasing temperature, in agreement with the increase of the spectral weight, see Fig.\ \ref{fig:sigma_t2g}. In contrast, the peak height for $i$\,=\,1 and 5 decreases from low temperature up to the phase transition temperature 77\,K, see Figs.\ 
\ref{fig:sigma_t2g} and \ref{fig:height}. This points towards a change of the linewidth.

\subsection{Determination of electronic parameters}
\label{sec:theo}

Based on the results from RIXS and optical spectroscopy, we turn our focus to the quantitative 
analysis of the multiplet energies and the effect of spin-orbit coupling on the electronic 
ground state. Considering a single site in cubic symmetry, the electronic energy levels are 
determined by the cubic crystal-field splitting 10\,$Dq$, the interelectronic Coulomb 
interaction captured by the Slater integrals $F^2$ and $F^4$, and spin-orbit coupling $\zeta$. 
The Slater integrals can also be expressed in terms of Hund’s coupling
$J_{\rm H}$\,=\,$1/14 (F^2+F^4)$ within the entire $5d$ shell \cite{Georges13}. 
Using \textsc{Quanty} \cite{Haverkort12, Haverkort16}, we calculate the effect of these 
parameters on the $5d^3$ energy levels, see Fig.\ \ref{fig:energy}. 
As mentioned in Sect.\ \ref{sec:assign}, $J_{\rm H}$ splits the 20 $t_{2g}^3$ states into 
the cubic multiplets $^4A_2$, $^2E$, $^2T_1$ and $^2T_2$, while $\zeta$ gives rise to a further 
lifting of degeneracies, yielding the terms $\Gamma_8$ (4 times),  $\Gamma_6$,  and $\Gamma_7$.

\subsubsection{$t_{2g}$-only limit for 10\,$Dq$\,=\,$\infty$}
\label{sec:Kanamori}

\begin{figure}[t]
	\centering
	\includegraphics[width=\columnwidth]{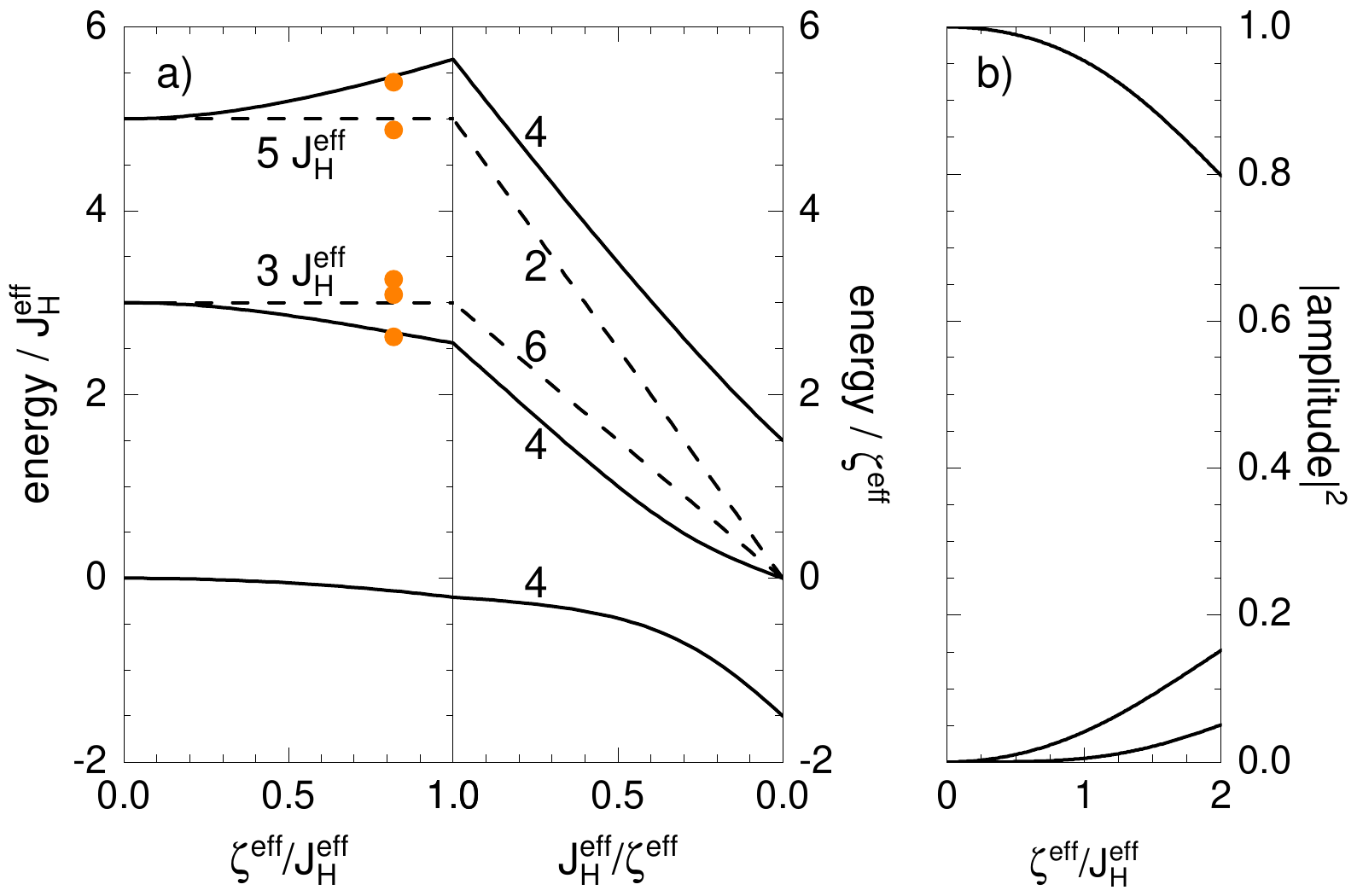}
	\caption{{\bf Absolute energies of the $t_{2g}^3$ states and ground state wavefunction 
	for 10\,$Dq$\,=\,$\infty$.} 
	a) To obtain excitation energies, the ground state energy 
	$E_0$ (lowest line) has to be subtracted, see Eq.\ \eqref{eq:eigenvalues}.
	For $\zeta^{\rm eff}$\,=\,0, the ground state is at zero while the excited states 
	are at 3\,$J_{\rm H}^{\rm eff}$ and 5\,$J_{\rm H}^{\rm eff}$. 
	Spin-orbit coupling mixes some of the states (solid lines) while other states 
	remain unaffected (dashed), giving rise to, in total, four excitation energies. 
	The best agreement with experiment (symbols) is obtained for 
	$\zeta^{\rm eff}/J_{\rm H}^{\rm eff}$\,=\,0.82 with 
	$J_{\rm H}^{\rm eff}$\,=\,343\,meV.\@
	Numbers in the right panel denote the number of states. 
	b) Squared amplitudes of the three contributions to the ground state wavefunction 
	corresponding to the solid lines in a). 
	}
	\label{fig:Kanamori}
\end{figure}

We start the quantitative analysis within the $t_{2g}$-only Kanamori picture for 10\,$Dq$\,=\,$\infty$, neglecting any admixture of $e_g$ states. This approximation is often 
chosen in theory, for instance in the discussion of the possible role of the Jahn-Teller 
effect \cite{Streltsov20}. The matrix elements describing the mixing of the cubic 
$t_{2g}^3$ multiplets due to spin-orbit coupling have been reported in, e.g., 
Refs.\ \cite{Hoggard81,Kamimura60}, 
reducing the problem to two $5\times 5$ matrices that we address in 
Appendix B.\@ 
The $t_{2g}$-only picture shows four excitation energies. As discussed above, 
the degeneracy between 
the $^2E$ and $^2T_1$ states is not lifted for 10\,$Dq$\,=\,$\infty$ and $\zeta$\,=\,0.  
Figure \ref{fig:Kanamori} plots the absolute energies, covering both limits, $LS$ coupling 
for $\zeta^{\rm eff}/J_{\rm H}^{\rm eff}$\,$\rightarrow$\,0 and $jj$ coupling for 
$\zeta^{\rm eff}/J_{\rm H}^{\rm eff}$\,$\rightarrow$\,$\infty$. 
Three of the absolute energies (solid lines in Fig.\ \ref{fig:Kanamori}a) are given by
\begin{eqnarray}
	\label{eq:eigenvalues}
	\nonumber
	\frac{E_n}{J_{\rm H}^{\rm eff}} & = & \frac{3+5}{3}-\frac{\tau}{3} \\
	&\cdot & \cos\left[  \frac{1}{3} \arccos\left( \frac{4(284-3\tau^2)}{\tau^{3}}\right)+2\pi \frac{n}{3} \right]
\end{eqnarray}
with $n$\,=\,0, 1, and 2 and
\begin{equation}
\tau= 2 \sqrt{3^2 + 5^2-3\cdot 5 + 3\left( \frac{3}{2}\frac{\zeta^{\rm eff}}{J_{\rm H}^{\rm eff}} \right)^2 } \, .
\end{equation}
The expression of $\tau$ reflects the energies in the limits 
$\zeta^{\rm eff}/J_{\rm H}^{\rm eff}$\,$\rightarrow$\,0 
and  $\zeta^{\rm eff}/J_{\rm H}^{\rm eff}$\,$\rightarrow$\,$\infty$. 
Two further absolute energies equal $3 J_{\rm H}^{\rm eff}$ and $5 J_{\rm H}^{\rm eff}$ 
(dashed lines), unaffected by spin-orbit coupling. 
Note that $E_1\! > \! E_2$ and that the 
excitation energies are obtained by subtracting the ground state energy $E_0$. 
The states at $3 J_{\rm H}^{\rm eff}$ and $5 J_{\rm H}^{\rm eff}$ (dashed lines), not mixing with the $S$\,=\,3/2 $^4A_2$ multiplet, correspond to the absorption bands with narrow features, $i$\,=\,2-4, while the states with $n$\,=\,1 and 2 (solid lines) can be identified with the broader bands for $i$\,=\,1 and 5 in the optical conductivity.

The best description of the five experimental intra-$t_{2g}$ energies $E_{0,i}$ 
is achieved for $\zeta^{\rm eff}/J_{\rm H}^{\rm eff}$\,$\approx$\,0.8 with 
$J_{\rm H}^{\rm eff}$\,=\,343\,meV, see symbols in Fig.\ \ref{fig:Kanamori}a).
This ratio falls within the range of intermediate coupling. It is about a factor 
of two too small to cause a pronounced Jahn-Teller effect, 
see Fig.\ \ref{fig:structure}.

The description of the experimental peak energies is reasonable but not excellent, 
as illustrated in Fig.\ \ref{fig:Kanamori}a) and by the following example. The energy scale $J_{\rm H}^{\rm eff}$\,$=$\,343\,meV is given by 
$5J_{\rm H}^{\rm eff}$\,$\approx$\,$E_{0,4}$\,=\,1723\,meV.\@ 
This predicts a splitting of about 0.7\,eV between the states at roughly 
$5J_{\rm H}^{\rm eff}$ and $3J_{\rm H}^{\rm eff}$, which is substantially larger 
than the corresponding experimental values 
$E_{0,4}-E_{0.3}$\,=\,(1723-1164)\,meV and $E_{0,4}-E_{0.2}$\,=\,(1723-1107)\,meV.\@ 
To achieve a better description of the experimental data, the effect of higher lying 
states has to be taken into account.

\subsubsection{Parameters for the entire $d$ shell}

Going beyond the $t_{2g}$-only model, we consider the entire $5d$ shell and 
fit the five intra-$t_{2g}$ energies $E_{0,i}$ 
taken from $\sigma_1(\omega)$ and the energy of the RIXS peak at 3.3\,eV.\@ 
Using the widely employed value $F^4/F^2$\,=\,36/55 (equivalent to Racah $C/B$\,=\,4), 
the best agreement is found for 
10\,$Dq$\,=\,3.25\,eV, $\zeta$\,=\,290\,meV, and $F^2$\,=\,3.93\,eV.\@
This yields $J_{\mathrm{H}}$\,=\,464\,meV and 
$\zeta/J_{\mathrm{H}}$\,$\approx$\,0.6, 
i.e., intermediate coupling. Note that we can also obtain 
$J_{\rm H}^{\rm eff}$\,=\,$(3/49)F^2+ (20/441)F^4$\,$\approx$\,0.36\,eV 
from these parameters, in good agreement with the analysis described above. 
The fit yields $E_{0,i}^\mathrm{fit}$\,=\,945, 1064, 1132, 1773, and 1901 \,meV  
for the intra-$t_{2g}$ excitations and 3.31\,eV for the $t_{2g}$-to-$e_g$ transition, 
see symbols in Fig.\ \ref{fig:energy}.

While $E_{0,1}$ and $E_{0,5}$ are described within 3\,meV, the fit underestimates 
$E_{0,2}$ and $E_{0,3}$ but overestimates $E_{0,4}$ by about 3-4\,\%.  
This corresponds to the difficulty discussed for the $t_{2g}$-only model concerning 
the comparably small experimental splitting between 
the states attributed to about $5J_{\rm H}^{\rm eff}$ and $3J_{\rm H}^{\rm eff}$. 
A decrease of 10\,$Dq$ indeed lowers $E_{0,4}$ more strongly than $E_{0,3}$ and $E_{0,2}$,  
i.e., the admixture of $e_g$ states reduces the splitting but does not yet yield perfect agreement for the appropriate value of 10\,$Dq$ that is fixed by the RIXS data. 
The description of the experimental peak energies can be further improved by, e.g., 
considering the ratio $F^4/F^2$ as a fit parameter or by allowing for charge-transfer processes to 
the ligands, which adds further fit parameters. We refrain from following this path since it has no profound impact on the result for $\zeta/J_{\rm H}$.

\subsubsection{Effect of spin-orbit coupling on the ground state}

Beyond spin-orbit coupling or $\zeta/J_{\rm H}$, the effective magnetic 
moment of the $5d^3$ configuration is affected by the cubic crystal-field splitting, 
by deviations from cubic symmetry, and by exchange interactions \cite{Figgis61}. 
The latter can be expected to be particularly relevant in K$_2$ReCl$_6$ due to the strong 
exchange frustration of the \textit{fcc} lattice \cite{Revelli19a}. 
However, exchange interactions will be rather small due to the large Re-Re distance. 
The admixture of $e_g$ states for finite 10\,$Dq$ reduces the effective moment $2\sqrt{S(S+1)}$\,$\mu_B$ 
for $\zeta$\,=\,0 approximately by the factor $1-(4/3)\zeta/10\,Dq$ \cite{Hitchman}, 
which in K$_2$ReCl$_6$ equals 0.88 according 
to our results. In the following, we focus on the effect of spin-orbit coupling.

For $\zeta$\,=\,0, the ground state is given by the $S$\,=\,3/2 $^4\!A_2$ multiplet with quenched orbital moment. In this state, each $t_{2g}$ orbital is occupied by one electron. It is not split 
by non-cubic distortions. 
Finite spin-orbit coupling causes an admixture of higher-lying states, in particular of the $^2T_2$ multiplet, the one highest in energy, see Fig.\ \ref{fig:energy}. 
This adds orbital moment and drives the Jahn-Teller activity. 
For an intuitive picture, we employ the $t_{2g}$-only Kanamori model.

The five energies have been described in Sect.\ \ref{sec:Kanamori}, and the eigenstates and expectation values of $L_z$ are discussed in Appendix B.\@ 
The ground state exhibits contributions from all four cubic $t_{2g}^3$ multiplets but can be written  
as a superposition of three terms, and their weights or squared amplitudes are depicted 
in Fig.\ \ref{fig:Kanamori}b). 
To first order, the weights of the admixed states increase like 
(1/25)\,$(\zeta^{\rm eff}/J_{\rm H}^{\rm eff})^2$ and (1/180)\,$(\zeta^{\rm eff}/J_{\rm H}^{\rm eff})^4$.
For $\zeta^{\rm eff}/J_{\rm H}^{\rm eff}$\,=\,0.8, as we find for K$_2$ReCl$_6$, more than 97\,\% of the weight is still contributed by the $^4\!A_2$ multiplet. 
If we include the $e_g$ states and consider the parameters obtained from the fit 
discussed above, we still find 93\,\% of the ground state weight to be carried by the 
$S$\,=\,3/2 multiplet.

We focus on the state showing $S_z$\,=\,3/2 for $\zeta$\,=\,0. For finite $\zeta$, we find 
for the expectation values of $L_z$ and $J_z$\,=\,$S_z - L_z$ in leading order 
\begin{align}
	\langle L_z \rangle &\approx -\frac{1}{25}\left(\zeta^{\rm eff}/J_{\rm H}^{\rm eff} \right)^2 
	\\
	\langle J_z \rangle &\approx \frac{3}{2} 
	- \frac{2}{15^2}\left(\zeta^{\rm eff}/J_{\rm H}^{\rm eff}\right)^4 \, .
\end{align}
The finite expectation value of $L_z$ predominantly arises from the admixture of the $^2T_2$ multiplet into the ground state, and this multiplet is Jahn-Teller active. 
However, for $\zeta^{\rm eff}/J_{\rm H}^{\rm eff}$\,=\,0.8, we find 
$\langle L_z \rangle$\,$\approx$\,0.027, a small value. 
In $LS$ coupling, $S_z$ and $L_z$ sum up to 3/2. The deviations from this value describe the gradual transition to $jj$ coupling. This deviation increases slowly in fourth order in 
$\zeta^{\rm eff}/J_{\rm H}^{\rm eff}$. 
For $\zeta^{\rm eff}/J_{\rm H}^{\rm eff}$\,=\,0.8, 
we find $\langle J_z \rangle$\,$\approx$\,1.497, very close to 3/2.  
This agrees with the result of a Curie-Weiss fit of the magnetic susceptibility  
that finds an effective magnetic moment very close to the value expected for a 
$S$\,=\,3/2 compound \cite{Bertin22}. 
Note, however, that our analysis is restricted to the Kanamori model. 
Inspection of the energies in Fig.\ \ref{fig:Kanamori} shows that a more pronounced change of the ground state character occurs for 
$\zeta^{\rm eff}/J_{\rm H}^{\rm eff}$\,$\approx$\,2. Note that this agrees with the result of Streltsov and Khomskii \cite{Streltsov20} for the range where a strong Jahn-Teller effect sets in, 
see Fig.\ \ref{fig:structure}b).

\section{Conclusion}

For the half-filled $t_{2g}$ shell of K$_2$ReCl$_6$, the effect of spin-orbit coupling is 
more subtle than for $5d$ transition-metal compounds with other electron configurations.  
We studied the electronic excitations with RIXS and optical spectroscopy. 
In the optical conductivity $\sigma_1(\omega)$, the narrow, phonon-assisted intra-$t_{2g}$ 
features qualify as text book examples of on-site $d$-$d$ excitations. 
We find that spin-orbit coupling is sizable, $\zeta$\,=\,0.29\,eV.\@  
In the often employed $t_{2g}$-only Kanamori model, spin-orbit coupling causes 
an admixture of mainly $^2T_2$ character into the ground state. 
The corresponding orbital moment opens the door for Jahn-Teller activity, 
but the admixture in leading order increases only quadratically, 
$\propto (\zeta/J_{\rm H})^2$, and $\zeta/J_{\rm H}$\,=\,0.6 is too small to drive a 
sizable Jahn-Teller distortion. 
The $S$\,=\,3/2 multiplet carries about 97\,\% of the ground state weight 
in the Kanamori picture. Additionally taking into account the $e_g$ states, 
we find that 93\,\% of the weight stems from the $S$\,=\,3/2 multiplet. 
However, spin-orbit coupling may still leave its fingerprints, for instance causing 
anisotropy gaps in the magnon dispersion as discussed for 
$5d^3$ osmates \cite{Kermarec15,Calder16,Taylor16,Taylor18} 
or the pronounced magneto-elastic effects reported for K$_2$ReCl$_6$ \cite{Bertin22}. 
Concerning structural changes, we could not resolve a non-cubic splitting of the on-site 
$d$-$d$ excitations above $T_N$, the non-cubic crystal-field hence has to be small. 
However, we find subtle differences in the line shape comparing the excitations to 
quartets that are expected to split with the Kramers doublets that are insensitive 
to a non-cubic crystal field. The former show broader features and a stronger coupling 
to the lattice. Moreover, our careful analysis of the optical data provides evidence 
for a lowering of the ground state energy by about 3\,meV from the highest phase 
transition temperature 111\,K down to low temperature. 
The question whether the Jahn-Teller effect plays any role in the structural phase transitions 
will have to be addressed by thorough structural studies. Still, our results firmly establish 
that such effects can only be small. 

\begin{figure}[b]
	\centering
	\includegraphics[width=0.98\columnwidth]{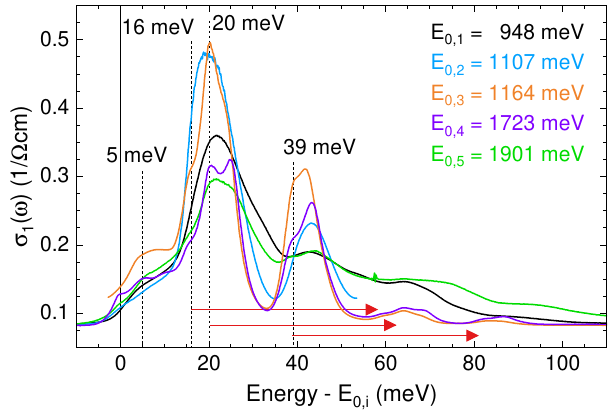}
	\includegraphics[width=0.98\columnwidth]{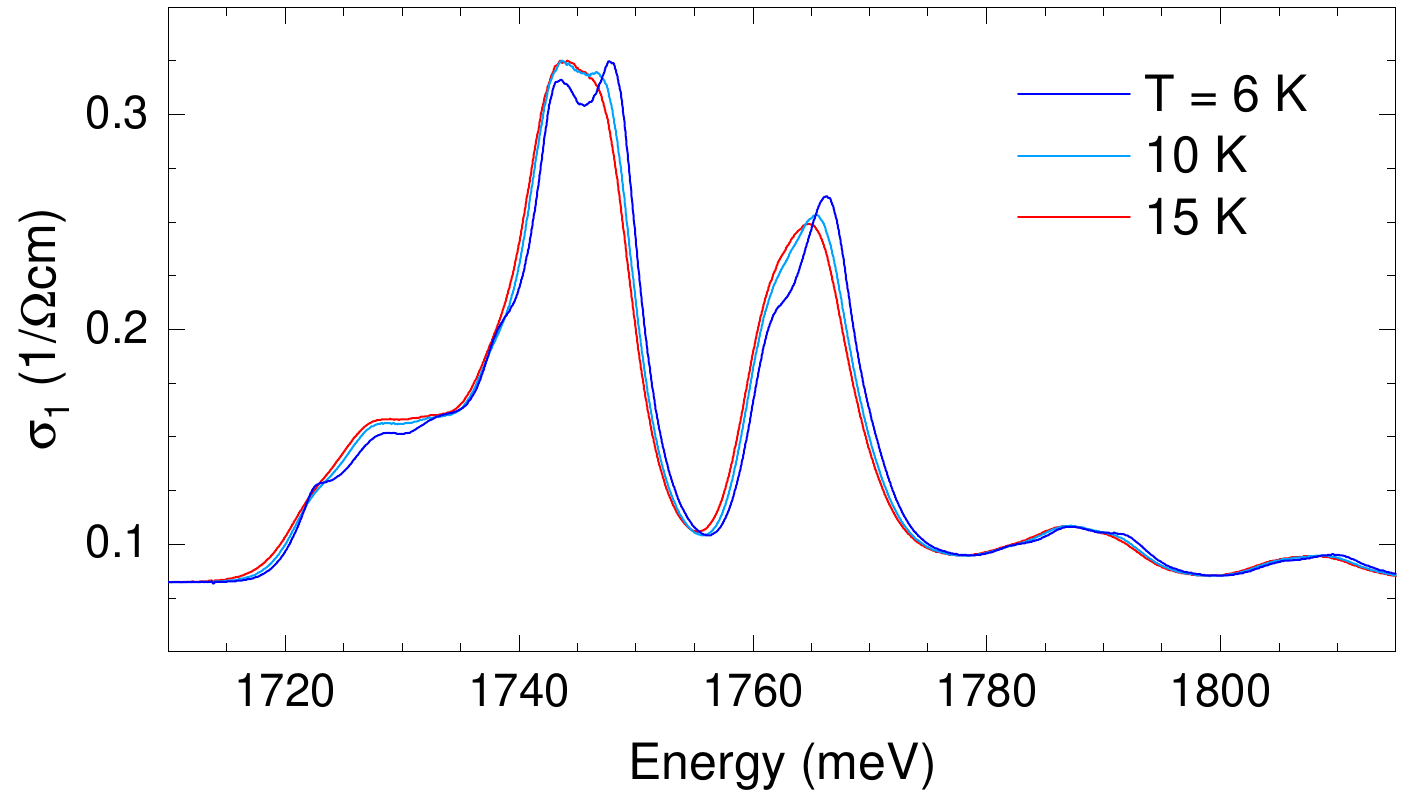}
	\caption{{\bf Peak structure of intra-$t_{2g}$ excitations below $T_N$.} 
		Top: Comparison of the line shape of the five absorption bands at 6\,K.\@ 
		Bottom: Temperature dependence of the band at $E_{0,4}$\,=\,1723\,meV.\@
	}
	\label{fig:shift6K}
\end{figure}

\appendix

\section{Line shape below $T_N$}
\label{sect:AppA}

In a state with long-range antiferromagnetic order, the spin-forbidden on-site 
$d$-$d$ excitations may show magnon sidebands. The joint excitation carries 
$\Delta S$\,=\,0 and does not require to involve the excitation of a phonon, 
as discussed for, e.g., compounds with $3d^3$ Cr$^{3+}$ ions \cite{Schmidt13}.  
However, magnon sidebands have also been reported for phonon-assisted features, both 
in Cr compounds and in K$_2$ReCl$_6$ \cite{Schmidt13,Szymczak80,Eremin11,Bettinelli88}. 
Below $T_N$\,=\,12\,K, we observe a splitting of some of the absorption peaks, and 
this splitting is most pronounced for the band with $E_{0,4}$\,=\,1723\,meV, 
see bottom panel of Fig.\ \ref{fig:shift6K}. 
A comparison of the line shape of the five absorption bands as a function of 
$E\! - \! E_{0,i}$ at 6\,K is given in the top panel of Fig.\ \ref{fig:shift6K}.

\section{Eigenstates and $L_z$ in Kanamori model}

We stick to the $t_{2g}$-only Kanamori model for analytic expressions of the energies and 
wavefunctions of the $t_{2g}^3$ states. For $\zeta$\,=\,0, Coulomb interactions yield 
the cubic multiplets $^4\!A_2$, $^2E$, $^2T_1$, and $^2T_2$. The corresponding eigenstates 
are given below in Sect.\ \ref{sec:eigenzeta0}, 
where we use $^2E(u)$ and $^2E(v)$ and, e.g., $^2T_1(u)$, $^2T_1(v)$, and $^2T_1(w)$ to distinguish states with orbital degeneracy. 
The multiplets are mixed by spin-orbit coupling
\begin{equation}
	H_{ls} = \zeta \, \vec{l}\vec{s} \, ,  
\end{equation}
where $\vec{l}$ and $\vec{s}$ denote the orbital and spin angular momenta of an individual 
electron. 
The problem decouples into two $5\times 5$ matrices \cite{Hoggard81}.
Using $x$\,=\,$\zeta^{\rm eff}/J_{\rm H}^{\rm eff}$, the matrix for the  basis states 
$^4A_2(\mp \frac{1}{2})$, $^2E(v,\mp \frac{1}{2})$, $^2T_2(v,\mp \frac{1}{2})$, $^2T_1(w,\pm \frac{1}{2})$, and $^2T_2(w,\pm \frac{1}{2})$ (see Sect.\ \ref{sec:eigenzeta0})
is given by 
\begin{equation}
	\label{matrixA2} H_1 =  J_{\rm H}^{\rm eff}
	\left( \begin{array}{ccccc}
		0 & 0 & 2x/\sqrt{6}  & 0 & x/\sqrt{3} \\ 
		0 & 3 & -x/\sqrt{3} & 0 & -x/\sqrt{6}\\ 
		2x/\sqrt{6}  & -x/\sqrt{3} & 5 &   -x/\sqrt{2} & 0 \\ 
		0  & 0 &  -x/\sqrt{2} & 3  & -x/2 \\ 
		x/\sqrt{3}  & -x/\sqrt{6} & 0 & -x/2  & 5  
	\end{array}  \right) 
\end{equation}
while for the basis states $^4A_2(\mp \frac{3}{2})$, $^2E(u,\pm \frac{1}{2})$, $^2T_1(u,\pm\frac{1}{2})$, $^2T_1(v,\mp\frac{1}{2})$, and $^2T_2(u,\mp \frac{1}{2})$ it becomes
\begin{equation}
	\label{matrixA1} H_2 =  J_{\rm H}^{\rm eff}
	\left( \begin{array}{ccccc}
		0 & 0 & 0  & 0 & x\\ 
		0 & 3 & 0 & 0 & -x/\sqrt{2}\\ 
		0  & 0 & 3 &   0 & x/\sqrt{2} \\ 
		0  & 0 &  0 & 3  & x/2 \\ 
		x & -x/\sqrt{2} & x/\sqrt{2} & x/2  & 5  
	\end{array}  \right) \, .
\end{equation}
We will discuss $H_1$ first. The analysis of $H_2$ is very similar.

\subsection{Solutions of $H_1$ for finite spin-orbit coupling}

Among the five eigenstates in the subspace of $H_1$, there are two states with constant energies $E(\Gamma_8^\prime)$\,=\,$3 J_{\rm H}^{\rm eff}$ 
and $E(\Gamma_7)$\,=\,$5 J_{\rm H}^{\rm eff}$,
see Fig.\ \ref{fig:Kanamori}. These are given by
\begin{eqnarray}
	\nonumber
    |\Gamma_8^\prime\rangle_1
	& = & -\sqrt{\frac{3}{5}}\, | ^2E(v,\mp \frac{1}{2})\rangle   + \sqrt{\frac{2}{5}}\, | ^2T_1(w,\pm \frac{1}{2})\rangle \\
	|\Gamma_7 \rangle 
	& = & -\sqrt{\frac{1}{3}}\, | ^2T_2(v,\mp \frac{1}{2})\rangle   + \sqrt{\frac{2}{3}}\, | ^2T_2(w,\pm \frac{1}{2}) \rangle \, . 
\end{eqnarray}
Additionally, we consider the states 
\begin{eqnarray}
	\nonumber
	|a \rangle & = &  | ^4A_2(\mp \frac{1}{2})\rangle \\ \nonumber
	|b \rangle & = & \sqrt{\frac{2}{5}}\, | ^2E(v,\mp \frac{1}{2})\rangle 
	 + \sqrt{\frac{3}{5}}\, | ^2T_1(w,\pm \frac{1}{2})\rangle \\
	|c \rangle & = & \sqrt{\frac{2}{3}}\, | ^2T_2(v,\mp \frac{1}{2})\rangle   +\sqrt{\frac{1}{3}}\, | ^2T_2(w,\pm \frac{1}{2}) \rangle 
\end{eqnarray}
based on which the problem further reduces to the following Hamiltonian matrix 
that describes the mixing of $\Gamma_8$ states, 
\begin{equation}
	\label{matrixA3} H' =  J_{\rm H}^{\rm eff}
	\left( \begin{array}{ccc}
		0 & 0  & x  \\ 
		0 & 3 & -\sqrt{5}x/2  \\ 
		x & -\sqrt{5}x/2 & 5   
	\end{array}  \right) \, .
\end{equation}
With $n$\,=\,0, 1, and 2 and the eigenvalues $E_n$ given in Eq.\ (\ref{eq:eigenvalues}), 
the three eigenstates of $H'$ are described by
\begin{equation}\label{eq:A3-evalue}
	|n\rangle = \frac{1}{\alpha_n} \left(1, \ \frac{\sqrt{5}}{2} \frac{E_n}{3 J_{\rm H}^{\rm eff} - E_n}, \frac{E_n}{\zeta^{\rm eff}} \right) \, , 
\end{equation}
\begin{equation}\label{eq:A3-estate}
	\alpha_n = \sqrt{ 1 + \frac{5}{4} \left(\frac{E_n}{3 J_{\rm H}^{\rm eff} - E_n}\right)^2 + \left(\frac{E_n}{\zeta^{\rm eff}}\right)^2 } \, . 
\end{equation}
The ground state, or more precisely one doublet of the $\Gamma_8$ ground state quartet, explicitly reads
\begin{equation}\label{eq:first_gs_doublet}
	|0\rangle_{\mp \frac{1}{2}} = \frac{1}{\alpha_0} | ^4A_2(\mp \frac{1}{2})\rangle 
	+ \frac{\sqrt{5}}{2\alpha_0} \frac{E_0}{3 J_{\rm H}^{\rm eff} - E_0} |b \rangle 
	+ \frac{E_0}{\alpha_0\,\zeta^{\rm eff}} |c \rangle \, .
\end{equation}
For $\zeta^{\rm eff}/J_{\rm H}^{\rm eff}$\,=\,0.8, appropriate for K$_2$ReCl$_6$, 
the three amplitudes are 0.9858, -0.0462, and -0.1617. 
The expectation value for $L_z$ in $|0\rangle_{\mp \frac{1}{2}}$ amounts to 
\begin{align}
	\langle L_z \rangle_{\mp \frac{1}{2}} =\frac{3}{5} \left(\frac{\sqrt{5} E_0}{2 \alpha_0 (3J_{\rm H}^{\rm eff}-E_0)}\right)^2 +\frac{1}{3}\left(\frac{E_0}{\alpha_0 \zeta^{\rm eff}}\right)^2 \,.
\end{align}
In leading order, the ground state energy $E_0$ for small $\zeta^{\rm eff}/J_{\rm H}^{\rm eff}$ 
is given by 
\begin{equation}\label{eq:energy_approx}
	E_0 = -\frac{1}{5}\frac{(\zeta^{\rm eff})^2}{J_{\rm H}^{\rm eff}} \, , 
\end{equation}
which gives the approximation 
\begin{eqnarray}
	|0\rangle_{\mp \frac{1}{2}}  & \approx & \frac{1}{\beta_0} \left[ 
	| ^4A_2(\mp \frac{1}{2})\rangle  \right.
	\\ 	\nonumber
	& - & \left. \frac{\zeta^{\rm eff}}{5J_{\rm H}^{\rm eff}} 
	\left(  \sqrt{\frac{2}{3}}\, | ^2T_2(v,\mp \frac{1}{2})\rangle   +\sqrt{\frac{1}{3}}\, 
	| ^2T_2(w,\pm \frac{1}{2}) \rangle 
	\right) \right]
\end{eqnarray}
with $\beta_0$\,=\,$\sqrt{1+(\zeta^{\rm eff}/5J_{\rm H}^{\rm eff})^2}$. 
The leading order contribution to $L_z$ in $|0\rangle_{\mp \frac{1}{2}}$ hence reads
\begin{align}
	\langle L_z \rangle_{\mp \frac{1}{2}} & \approx \frac{1}{75}\left( 
	\frac{\zeta^{\rm eff}}{J_{\rm H}^{\rm eff}}\right)^2 \,. 
\end{align}

\subsection{Solutions of $H_2$ for finite spin-orbit coupling}

Also in the subspace of $H_2$ there are two states with constant energies 
$E(\Gamma_6)$\,=\,$E(\Gamma_8^\prime)$\,=\,$3 J_{\rm H}^{\rm eff}$,
see Fig.\ \ref{fig:Kanamori}. 
These read
\begin{align}
	|\Gamma_6\rangle
	& =\sqrt{\frac{1}{3}} |^2T_1(u,\pm \frac{1}{2})\rangle  - 
	\sqrt{\frac{2}{3}} |^2T_1(v,\mp \frac{1}{2})\rangle    \nonumber\\
	|\Gamma_8^\prime\rangle_2 
	&= \sqrt{\frac{3}{5}} |^2 E(u,\pm \frac{1}{2})\rangle +
	   \sqrt{\frac{4}{15}}|^2T_1(u,\pm \frac{1}{2})\rangle  \nonumber\\  
     & + \sqrt{\frac{2}{15}}|^2T_1(v,\mp \frac{1}{2})\rangle . 
\end{align}
Choosing the remaining states as 
\begin{align}\label{eq:A3_three-states}
	|a^\prime\rangle & =|^4A_2(\mp \frac{3}{2})\rangle 
	\nonumber\\
	|b^\prime \rangle &= \sqrt{\frac {2}{5}} \left(|^2 E(u,\pm \frac{1}{2})\rangle-|^2T_1(u,\pm \frac{1}{2})\rangle \right. \nonumber\\
	& \left. -\frac{1}{\sqrt{2}}|^2T_1(v,\mp \frac{1}{2})|\rangle\right) \nonumber\\
	|c^\prime\rangle &=  |^2T_2(u,\mp \frac{1}{2})\rangle , 
\end{align}
the problem reduces to the same 3x3 matrix as earlier, i.e., Eq.\ \eqref{matrixA3}.
In particular, the eigenvalues and eigenstates are given by Eqs.\ 
\eqref{eq:eigenvalues} and \eqref{eq:A3-evalue}, respectively, but this time with the basis states $|a^\prime\rangle$, $|b^\prime\rangle$, and $|c^\prime\rangle$. 
The second doublet of the $\Gamma_8$ ground state manifold hence reads
\begin{equation}\label{eq:second_gs_doublet}
	|0\rangle_{\mp \frac{3}{2}} = \frac{1}{\alpha_0} | ^4A_2(\mp \frac{3}{2})\rangle 
	+ \frac{\sqrt{5}}{2\alpha_0} \frac{E_0}{3 J_{\rm H}^{\rm eff} - E_0} |b^\prime \rangle 
	+ \frac{E_0}{\alpha_0\,\zeta^{\rm eff}} |c^\prime \rangle \, . 
\end{equation}
The expectation value for $L_z$ in $|0\rangle_{\mp \frac{3}{2}}$ is given by
\begin{align}
	\langle L_z \rangle_{\mp \frac{3}{2}} =\frac{1}{5} \left(\frac{\sqrt{5} E_0}{2 \alpha_0 (3J_{\rm H}^{\rm eff}-E_0)}\right)^2 +\left(\frac{E_0}{\alpha_0 \zeta^{\rm eff}}\right)^2 \,.
\end{align}
Using the leading order approximation of $E_0$ in Eq.\ \eqref{eq:energy_approx}, one finds that the ground state is approximated by 
\begin{align}
	|0\rangle_{\mp \frac{3}{2}}&=\frac{1}{\beta_0}\left( | ^4A_2(\mp \frac{3}{2})\rangle 
	- \frac{\zeta^{\rm eff}}{5 J_{\rm H}^{\rm eff}} |^2T_2(u,\mp \frac{1}{2})\rangle \right) 
\end{align}
with, as above, $\beta_0$\,=\,$\sqrt{1+(\zeta^{\rm eff}/5J_{\rm H}^{\rm eff})^2}$. 
The $L_z$ expectation value is to leading order given by 
\begin{align}
	\langle L_z \rangle_{\mp \frac{3}{2}} &\approx \frac{1}{25}\left(\frac{\zeta^{\rm eff}}{J_{\rm H}^{\rm eff}}\right)^2.
\end{align}
The $^2T_2$ multiplet is Jahn-Teller active. Spin-orbit coupling causes a mixing in of 
$^2T_2$ character into the ground state, driving it Jahn-Teller active. 
In the Kanamori picture for $\zeta^{\rm eff}/J_{\rm H}^{\rm eff}$\,=\,0.8, we find 
$\langle L_z \rangle_{\mp \frac{3}{2}}$\,$\approx$\,0.027, a small value.
\\

\subsection{Eigenstates for $\zeta$\,=\,0}
\label{sec:eigenzeta0}

The 20 $t_{2g}^3$ eigenstates for $\zeta$\,=\,0 read
\begin{eqnarray}
		|^4A_2(\mp \frac{3}{2}) \rangle&= & c^\dagger_{xy-\sigma}c^\dagger_{xz-\sigma}c^\dagger_{yz-\sigma} |\rm vac\rangle\nonumber\\
	\nonumber
	|^4\!A_2 (\mp \frac{1}{2})\rangle & = & \frac{1}{\sqrt{3}}\left(
	c^\dagger_{xy\sigma} c^\dagger_{xz-\sigma} c^\dagger_{yz-\sigma}
	+	c^\dagger_{xy-\sigma} c^\dagger_{xz\sigma} c^\dagger_{yz-\sigma} \right. \\
	\nonumber
	& +	& \left. c^\dagger_{xy-\sigma} c^\dagger_{xz-\sigma} c^\dagger_{yz\sigma}
	\right) |{\rm vac}\rangle 
	\\ 
	| ^2E(u, \pm \frac{1}{2})\rangle&=& \frac{1}{\sqrt{2}} \left(c^\dagger_{xz\sigma}c^\dagger_{yz-\sigma}-c^\dagger_{xz-\sigma}c^\dagger_{yz\sigma} \right)c^\dagger_{xy\sigma} |\rm vac\rangle \nonumber\\
		\nonumber
	|^2\!E (v,\mp \frac{1}{2})\rangle & = & \frac{1}{\sqrt{6}}\left(
	2 c^\dagger_{xy\sigma} c^\dagger_{xz-\sigma} c^\dagger_{yz-\sigma} 
	-	c^\dagger_{xy-\sigma} c^\dagger_{xz\sigma} c^\dagger_{yz-\sigma} \right. \\
	\nonumber
	& -	& \left. c^\dagger_{xy-\sigma} c^\dagger_{xz-\sigma} c^\dagger_{yz\sigma}
	\right) |{\rm vac}\rangle 
\end{eqnarray}
\begin{eqnarray}
	|^2T_1(u,\pm \frac{1}{2})\rangle&=& \frac{i}{\sqrt{2}}\left(c^\dagger_{xz\sigma}c^\dagger_{xz-\sigma}-c^\dagger_{yz\sigma}c^\dagger_{yz-\sigma}\right)c^\dagger_{xy\sigma}|\rm vac\rangle\nonumber\\
	|^2T_1(v,\mp \frac{1}{2})\rangle&=&\frac{1}{2}\left[ \left(c^\dagger_{yz\sigma}c^\dagger_{yz-\sigma}-c^\dagger_{xy\sigma}c^\dagger_{xy-\sigma}\right)c^\dagger_{xz-\sigma}\right.\nonumber\\
	& \mp & \left. i \left(c^\dagger_{xy\sigma}c^\dagger_{xy-\sigma}- c^\dagger_{xz\sigma}c^\dagger_{xz-\sigma}\right)c^\dagger_{yz-\sigma} \right]|\rm vac\rangle\nonumber\\
	\nonumber
	|^2\!T_1 (w,\pm \frac{1}{2})\rangle & = & \frac{1}{2}\left[\left(
	c^\dagger_{yz\sigma} c^\dagger_{yz-\sigma} 
	-	c^\dagger_{xy\sigma} c^\dagger_{xy-\sigma}  \right) c^\dagger_{xz\sigma} \right. \\
	\nonumber
	& \mp &i\left. \left(  c^\dagger_{xy\sigma} c^\dagger_{xy-\sigma} 
	- c^\dagger_{xz\sigma} c^\dagger_{xz-\sigma} 
	\right)c^\dagger_{yz\sigma} \right]|{\rm vac}\rangle 
\end{eqnarray}
\begin{eqnarray}
	|^2T_2(u,\mp \frac{1}{2})\rangle&=&\frac{1}{2}\left[ \left(c^\dagger_{yz\sigma}c^\dagger_{yz-\sigma}+c^\dagger_{xy\sigma}c^\dagger_{xy-\sigma}\right)c^\dagger_{xz-\sigma}\right.\nonumber\\
	&\pm& \left.  i \left(c^\dagger_{xy\sigma}c^\dagger_{xy-\sigma}+ c^\dagger_{xz\sigma}c^\dagger_{xz-\sigma}\right)c^\dagger_{yz-\sigma} \right]|\rm vac \rangle,\nonumber\\
	\nonumber
	|^2\!T_2 (v,\mp \frac{1}{2})\rangle & = & \frac{i}{\sqrt{2}}\left(
	c^\dagger_{xz\sigma} c^\dagger_{xz-\sigma} + c^\dagger_{yz\sigma} c^\dagger_{yz-\sigma} \right) c^\dagger_{xy-\sigma} 
	|{\rm vac}\rangle 
	\\
	\nonumber
	|^2\!T_2 (w,\pm \frac{1}{2})\rangle & = & \frac{1}{2}\left[
	\left(c^\dagger_{yz\sigma} c^\dagger_{yz-\sigma} 
	+	c^\dagger_{xy\sigma} c^\dagger_{xy-\sigma}\right)  c^\dagger_{xz\sigma}\right. \\
	\nonumber
	& \pm	& \left. i \left(c^\dagger_{xy\sigma} c^\dagger_{xy-\sigma} 
	+c^\dagger_{xz\sigma} c^\dagger_{xz-\sigma} \right)c^\dagger_{yz\sigma}
	\right] |{\rm vac}\rangle 
\end{eqnarray}
where, e.g., $c^\dagger_{xy\sigma}$ creates an electron with spin $\sigma$ in the $xy$ orbital 
and $|{\rm vac}\rangle$ denotes vacuum.

\begin{acknowledgments}
We gratefully acknowledge fruitful discussions with J. van den Brink. 
We thank the European Synchrotron Radiation Facility for providing 
beam time at ID20 and technical support. 
Furthermore, we acknowledge funding from the Deutsche Forschungsgemeinschaft 
(DFG, German Research Foundation) through Project No.\ 277146847 -- CRC 1238 
(projects A02, B02, B03). 
M.H.\@ acknowledges partial funding by the Knut and Alice Wallenberg Foundation 
as part of the Wallenberg Academy Fellows project.
\end{acknowledgments}


\begin{thebibliography}{99}   
	
\bibitem{WitczakKrempa14}
W. Witczak-Krempa, G. Chen, Y. B. Kim, and L. Balents, 
\textit{Correlated Quantum Phenomena in the Strong Spin-Orbit Regime}, 
Annu. Rev. Condens. Matter Phys. \textbf{5}, 57 (2014).

\bibitem{Rau16}
J. G. Rau, E. K.-H. Lee, and H.-Y. Kee, 
\textit{Spin-Orbit Physics Giving Rise to Novel Phases in Correlated Systems: 
	Iridates and Related Materials}, 
Annu. Rev. Condens. Matter Phys. \textbf{7}, 195 (2016). 

\bibitem{Schaffer16}
R. Schaffer, E. K.-H. Lee, B.-J. Yang, and Y. B. Kim, 
\textit{Recent progress on correlated electron systems with strong spin-orbit coupling}, 
Rep. Prog. Phys. \textbf{79}, 094504 (2016).

\bibitem{Streltsov20}
S. V. Streltsov  and D. I. Khomskii,
\textit{Jahn-Teller Effect and Spin-Orbit Coupling: Friends or Foes?},
Phys. Rev. X \textbf{10}, 031043 (2020).

\bibitem{Takayama21}
T. Takayama, J. Chaloupka, A. Smerald, G. Khaliullin, and H. Takagi, 
\textit{Spin–Orbit-Entangled Electronic Phases in $4d$ and $5d$ Transition-Metal Compounds}, 
J. Phys. Soc. Jpn. \textbf{90}, 062001 (2021).

\bibitem{Khomskii21}
D. I. Khomskii and S. V. Streltsov, 
\textit{Orbital Effects in Solids: Basics, Recent Progress, and Opportunities}, 
Chem. Rev.  \textbf{121}, 2992 (2021).	 

\bibitem{Jackeli09}
G. Jackeli and G. Khaliullin,
\textit{Mott Insulators in the Strong Spin-Orbit Coupling Limit: From Heisenberg to a 
	Quantum Compass and Kitaev Models},
Phys. Rev. Lett. \textbf{102}, 017205 (2009).

\bibitem{Winter17}
S. M. Winter, A. A. Tsirlin, M. Daghofer, J. van den Brink, Y. Singh, P. Gegenwart, 
and R. Valent{\'\i},
\textit{Models and materials for generalized Kitaev magnetism},
J. Phys.: Condens. Matter \textbf{29}, 493002 (2017).

\bibitem{Chun15}
S. H. Chun, J.-W. Kim, J. Kim, H. Zheng, C. C. Stoumpos, C. D. Malliakas, J. F. Mitchell, 
K. Mehlawat, Y. Singh, Y. Choi, T. Gog, A. Al-Zein, M. Moretti Sala, M. Krisch, J. Chaloupka, 
G. Jackeli, G. Khaliullin, and  B. J. Kim, 
\textit{Direct evidence for dominant bond-directional interactions in a honeycomb lattice 
iridate Na$_2$IrO$_3$}, 
Nat. Phys. \textbf{11}, 462 (2015).

\bibitem{Revelli19a}
A. Revelli, C.C. Loo, D. Kiese, P. Becker, T. Fr\"ohlich, T. Lorenz, M. Moretti Sala, 
G. Monaco, F.L. Buessen, J. Attig, M. Hermanns, S. V.~Streltsov, D. I. Khomskii, 
J. van den Brink, M. Braden, P.H.M. van Loosdrecht, S. Trebst, A. Paramekanti, 
and M. Gr\"{u}ninger,
\textit{Spin-orbit entangled $j$\,=\,$1/2$ moments in Ba$_2$CeIrO$_6$: A  frustrated 
	fcc quantum magnet},
Phys. Rev. B \textbf{100}, 085139 (2019).

\bibitem{Magnaterra23}
M. Magnaterra, K. Hopfer, Ch. J. Sahle, M. Moretti Sala, G. Monaco, J. Attig, 
C. Hickey, I.-M. Pietsch, F. Breitner, P. Gegenwart, M. H. Upton, Jungho Kim, S. Trebst, 
P. H. M. van Loosdrecht, J. van den Brink, and M. Gr\"{u}ninger, 
\textit{RIXS observation of bond-directional nearest-neighbor excitations 
	in the Kitaev material Na$_2$IrO$_3$}, 
arXiv:2301.08340


\bibitem{Chen10}
G. Chen, R. Pereira, and L. Balents, 
\textit{Exotic phases induced by strong spin-orbit coupling in ordered double perovskites}, 
Phys.
Rev. B \textbf{82}, 174440 (2010).

\bibitem{Natori16}
W. M. H. Natori, E. C. Andrade, E. Miranda, and R. G. Pereira,
\textit{Chiral spin-orbital liquids with nodal lines}, 
Phys. Rev. Lett.
\textbf{117}, 017204 (2016).


\bibitem{Romhanyi17}
J. Romhanyi, L. Balents, and G. Jackeli, 
\textit{Spin-orbit dimers and noncollinear phases in $d^1$ cubic double perovskites}, 
Phys. Rev.
Lett. \textbf{118}, 217202 (2017).


\bibitem{Ishikawa19}
H. Ishikawa, T. Takayama, R. K. Kremer, J. Nuss, R. Dinnebier, K. Kitagawa, K. Ishii, 
and H. Takagi, 
\textit{Ordering of hidden multipoles in spin-orbit entangled $5d^1$ Ta chlorides}, 
Phys. Rev. B \textbf{100}, 045142 (2019). 

\bibitem{Tehrani23}
A. Mansouri Tehrani, J.-R. Soh, J. P\'asztorov\'a, M. E. Merkel, I. \v{Z}ivkovi\'c, 
H. M. R\o nnow, and N. A. Spaldin, 
\textit{Charge multipole correlations and order in Cs$_2$TaCl$_6$}, 
Phys. Rev. Res. \textbf{5},
L012010 (2023).

\bibitem{Paramekanti20}
A. Paramekanti, D. D. Maharaj, and B. D. Gaulin, 
\textit{Octupolar order in $d$-orbital Mott insulators},
Phys. Rev. B \textbf{101}, 054439 (2020).

\bibitem{Lovesey20}
S. W. Lovesey and D. D. Khalyavin, 
\textit{Lone octupole and bulk magnetism in osmate $5d^2$ double perovskites},
Phys. Rev. B \textbf{102}, 064407 (2020).

\bibitem{Khaliullin21}
G. Khaliullin, D. Churchill, P. P. Stavropoulos, and H.-Y. Kee,
\textit{Exchange interactions, Jahn-Teller coupling, and multipole orders in 
	pseudospin one-half $5d^2$ Mott insulators}, 
Phys. Rev. Research \textbf{3}, 033163 (2021).

\bibitem{Voleti21}
S. Voleti, A. Haldar, and A. Paramekanti, 
\textit{Octupolar order and Ising quantum criticality tuned by strain and dimensionality:
	Application to $d$-orbital Mott insulators},
Phys. Rev. B \textbf{104}, 174431 (2021).

\bibitem{Pourovskii21}
L. V. Pourovskii, D. F. Mosca, and C. Franchini, 
\textit{Ferro-octupolar order and low-energy excitations in $d^2$ double perovskites of osmium}, 
Phys. Rev. Lett. \textbf{127}, 237201 (2021).

\bibitem{Rayyan23}
A. Rayyan, D. Churchill, and H.-Y. Kee,
\textit{Field-induced Kitaev multipolar liquid in spin-orbit coupled $d^2$ honeycomb Mott insulators}, 
Phys. Rev. B \textbf{107}, L020408 (2023).

\bibitem{Warzanowski23}
P. Warzanowski, M. Magnaterra, P. Stein, G. Schlicht, Q. Faure, Ch. J. Sahle, T. Lorenz, P. Becker, 
L. Bohat\'{y}, M. Moretti Sala, G. Monaco, P. H. M. van Loosdrecht, and M. Gr\"uninger, 
\textit{Electronic excitations in $5d^4$ $J$\,=\,0 Os$^{4+}$ halides studied by RIXS 
	and optical spectroscopy}, 
Phys. Rev. B \textbf{108}, 125120 (2023).

\bibitem{Yuan17}
B. Yuan, J. P. Clancy, A. M. Cook, C. M. Thompson, J. Greedan, G. Cao, B. C. Jeon, T. W. Noh, 
M. H. Upton, D. Casa, T. Gog, A. Paramekanti, and Y.-J. Kim, 
\textit{Determination of Hund's coupling in $5d$ oxides using resonant inelastic x-ray scattering},
Phys. Rev. B \textbf{95}, 235114 (2017).

\bibitem{Kusch18}
M. Kusch, V. M. Katukuri, N. A. Bogdanov, B. B\"{u}chner, T. Dey, D. V. Efremov, 
J. E. Hamann-Borrero, B. H. Kim, M. Krisch, A. Maljuk, M. Moretti Sala, S. Wurmehl, 
G. Aslan-Cansever, M. Sturza, L. Hozoi, J. van den Brink, and J. Geck, 
\textit{Observation of heavy spin-orbit excitons propagating in a nonmagnetic background:
	The case of (Ba,Sr)$_2$YIrO$_6$},
Phys. Rev. B \textbf{97}, 064421 (2018).

\bibitem{Nag18}
A. Nag, S. Bhowal, A. Chakraborty, M. M. Sala, A. Efimenko, F. Bert, P. K. Biswas, 
A. D. Hillier, M. Itoh, S. D. Kaushik, V. Siruguri, C. Meneghini, I. Dasgupta, and Sugata Ray, 
\textit{Origin of magnetic moments and presence of spin-orbit singlets in Ba$_2$YIrO$_6$},
Phys. Rev. B \textbf{98}, 014431 (2018).

\bibitem{Paramekanti18}
A. Paramekanti, D.\ J. Singh, B. Yuan, D. Casa, A. Said, Y.-J. Kim, and A.\ D. Christianson, 
\textit{Spin-orbit coupled systems in the atomic limit: rhenates, osmates, iridates},
Phys. Rev. B \textbf{97}, 235119 (2018).

\bibitem{Aczel22}
A. A. Aczel, Q. Chen, J. P. Clancy, C. dela Cruz, D. Reig-i-Plessis, 
G. J. MacDougall, C. J. Pollock, M. H. Upton, T. J. Williams, N. LaManna, 
J. P. Carlo, J. Beare, G. M. Luke, and H. D. Zhou, 
\textit{Spin-orbit coupling controlled ground states in the double perovskite
	iridates $A_2B$IrO$_6$ ($A$ = Ba, Sr; $B$ = Lu, Sc)}, 
Phys. Rev. Mater. \textbf{6}, 094409 (2022).

\bibitem{Fuchs18}
S. Fuchs, T. Dey, G. Aslan-Cansever, A. Maljuk, S. Wurmehl, B. B\"uchner, and V. Kataev, 
\textit{Unraveling the Nature of Magnetism of the $5d^4$ Double Perovskite Ba$_2$YIrO$_6$}, 
Phys. Rev. Lett. \textbf{120}, 237204 (2018).

\bibitem{Takahashi21}
H. Takahashi, H. Suzuki, J. Bertinshaw, S. Bette, C. M\"{u}hle, J. Nuss, R. Dinnebier, 
A. Yaresko, G. Khaliullin, H. Gretarsson, T. Takayama, H. Takagi, and B. Keimer, 
\textit{Nonmagnetic $J$\,=\,0 State and Spin-Orbit Excitations in K$_2$RuCl$_6$}, 
Phys. Rev. Lett. \textbf{127}, 227201 (2021).

\bibitem{Khaliullin13}
G. Khaliullin, 
\textit{Excitonic Magnetism in Van Vleck–type $d^4$ Mott Insulators}, 
Phys. Rev. Lett. \textbf{111}, 197201 (2013).

\bibitem{KhomskiiBook}
D. I. Khomskii, \textit{Transition metal compounds}, 
Cambridge University Press (2014).

\bibitem{Jain17}
A. Jain, M. Krautloher, J. Porras, G. H. Ryu, D. P. Chen, D. L. Abernathy, J. T. Park, 
A. Ivanov, J. Chaloupka, G. Khaliullin, B. Keimer, and B. J. Kim, 
\textit{Higgs mode and its decay in a two-dimensional antiferromagnet}, 
Nat. Phys. \textbf{13}, 633 (2017).

\bibitem{Kaushal21}
N. Kaushal, J. Herbrych, G. Alvarez, and E. Dagotto, 
\textit{Magnetization dynamics fingerprints of an excitonic condensate $t^4_{2g}$ magnet},
Phys. Rev. B \textbf{104}, 235135 (2021).

\bibitem{Taylor17}
A. E. Taylor, S. Calder, R. Morrow, H. L. Feng, M. H. Upton, M. D. Lumsden, K. Yamaura,
P. M. Woodward, and A. D. Christianson,
\textit{Spin-Orbit Coupling Controlled $J$\,=\,3/2 Electronic Ground State in $5d^3$ Oxides},
Phys. Rev. Lett. \textbf{118}, 207202 (2017).

\bibitem{Kermarec15}
E. Kermarrec, C. A. Marjerrison, C. M. Thompson, D. D. Maharaj, K. Levin, S. Kroeker, 
G. E. Granroth, R. Flacau, Z. Yamani, J. E. Greedan, and B. D. Gaulin, 
\textit{Frustrated fcc antiferromagnet Ba$_2$YOsO$_6$: Structural characterization, 
	magnetic properties, and neutron scattering studies}, 
Phys. Rev. B \textbf{91}, 075133 (2015).

\bibitem{Calder16}
S. Calder, J. G. Vale, N. A. Bogdanov, X. Liu, C. Donnerer, M. H. Upton, D. Casa, 
A. H. Said, M. D. Lumsden, Z. Zhao, J.-Q. Yan, D. Mandrus, S. Nishimoto, 
J. van den Brink, J. P. Hill, D. F. McMorrow, and A. D. Christianson, 
\textit{Spin-orbit-driven magnetic structure and excitation in the $5d$ pyrochlore Cd$_2$Os$_2$O$_7$},
Nat. Commun. \textbf{7}, 11651 (2016).

\bibitem{Taylor16}
A. E. Taylor, R. Morrow, R. S. Fishman, S. Calder, A. I. Kolesnikov, M. D. Lumsden, 
P. M. Woodward, and A. D. Christianson, 
\textit{Spin-orbit coupling controlled ground state in Sr$_2$ScOsO$_6$},
Phys. Rev. B \textbf{93}, 220408(R) (2016).

\bibitem{Taylor18}
A. E. Taylor, R. Morrow, M. D. Lumsden, S. Calder, M. H. Upton, A. I. Kolesnikov, M. B. Stone, 
R. S. Fishman, A. Paramekanti, P. M. Woodward, and A. D. Christianson,
\textit{Origin of magnetic excitation gap in double perovskite Sr$_2$FeOsO$_6$},
Phys. Rev. B \textbf{98}, 214422 (2018).

\bibitem{Calder17b}
S. Calder, D. J. Singh, V. O. Garlea, M. D. Lumsden, Y. G. Shi, K. Yamaura, 
and A. D. Christianson, 
\textit{Interplay of spin-orbit coupling and hybridization in Ca$_3$LiOsO$_6$ and Ca$_3$LiRuO$_6$},
Phys. Rev. B \textbf{96}, 184426 (2017).

\bibitem{Oleary70}
G. P. O'Leary and R. G. Wheeler,
\textit{Phase Transitions and Soft Librational Modes in Cubic Crystals},
Phys. Rev. B \textbf{1}, 4409 (1970).

\bibitem{Armstrong80}
R. L. Armstrong, 
\textit{Structural properties and lattice dynamics of $5d$ transition metal 
	antifluorite crystals}, 
Phys. Rep. \textbf{57}, 343 (1980).

\bibitem{Bertin22}
A. Bertin, T. Dey, D. Br\"{u}ning, D. Gorkov, K. Jenni, A. Krause, P. Becker, 
L. Bohat\'{y}, D. Khomskii, M. Braden, and T. Lorenz, 
\textit{Interplay of weak ferromagnetism, ferroelasticity and shape-memory effects 
	in the spin-orbit coupled antiferromagnet K$_2$ReCl$_6$}, 
arxiv:2207.11101 (2022).

\bibitem{Stein23}
P. Stein, T. C. Koethe, L. Bohat\'{y}, P. Becker, M. Gr\"uninger, 
and P. H. M. van Loosdrecht, 
\textit{Local symmetry breaking and low-energy continuum in K$_2$ReCl$_6$}, 
Phys. Rev. B \textbf{107}, 214301 (2023).

\bibitem{Haverkort12}
M. W. Haverkort, M. Zwierzcki, and O. K. Andersen, 
\textit{Multiplet ligand-field theory using Wannier orbitals}, 
Phys. Rev. B \textbf{85}, 165113 (2012).

\bibitem{Haverkort16}
M. W. Haverkort, 
\textit{Quanty for core level spectroscopy - excitons, resonances and band excitations 
	in time and frequency domain},
J. Phys.: Conf. Ser. \textbf{712}, 012001 (2016).

\bibitem{Busey62}
R. H. Busey and E. Sonder, 
\textit{Magnetic susceptibility of Potassium Hexachlororhenate (IV) and Potassium
	Hexabromorhenate (IV) from 5$^\circ$ to 300$^\circ$K}, 
J. Chem. Phys. \textbf{36}, 93 (1962).

\bibitem{Smith66}
H. G. Smith and G. E. Bacon, 
\textit{Neutron-diffraction study of magnetic ordering in K$_2$ReCl$_6$}, 
J. Appl. Phys. \textbf{37}, 979 (1966).

\bibitem{Smolentsev11}
N. Smolentsev, M. Sikora, A. V. Soldatov, K. O. Kvashnina, and P. Glatzel, 
\textit{Spin-orbit sensitive hard x-ray probe of the occupied and unoccupied $5d$ density of states}, 
Phys. Rev. B \textbf{84}, 235113 (2011).

\bibitem{Marcaud23}
G. Marcaud, A. T. Lee, A. J. Hauser, F. Y. Yang, S. Lee, D. Casa, M. Upton, T. Gog, 
K. Saritas, Y. Wang, M. P. M. Dean, H. Zhou, Z. Zhang, F. J. Walker, I. Jarrige, 
S. Ismail-Beigi, and Charles Ahn, 
\textit{Low-energy electronic interactions in ferrimagnetic Sr$_2$CrReO$_6$ thin films}, 
Phys. Rev. B \textbf{108}, 075132 (2023).

\bibitem{Frontini23}
F. I. Frontini, G. H.J. Johnstone, N. Iwahara, P. Bhattacharyya, N. A. Bogdanov, 
L. Hozoi, M. H. Upton, D. M. Casa, D. Hirai, and Y.-J. Kim,
\textit{Spin-orbit-lattice entangled state in $A_2$MgReO$_6$ ($A$\,=\,Ca, Sr, Ba) 
revealed by resonant inelastic X-ray scattering}, 
arxiv:2311.01621 (2023).

\bibitem{McMorrow02}
D. F. McMorrow, S. E. Nagler, K. A. McEwen, and S. D. Brown, 
\textit{Large enhancement of x-ray magnetic scattering at the $L$ edges of the 
	$5d$ transition metal antiferromagnet K$_2$ReCl$_6$},
J. Phys.: Condens. Matter \textbf{15}, L59 (2003).

\bibitem{Huotari2005}
S. Huotari, G. Vank\'{o}, F. Albergamo, C. Ponchut, H. Graafsma, 
C. Henriquet, R. Verbeni, and G. Monaco,
\textit{Improving the performance of high-resolution x-ray spectrometers with 
	position-sensitive pixel detectors}, 
J. Sync. Radiation \textbf{12}, 467 (2005).

\bibitem{Huotari2006}
S. Huotari, F. Albergamo, G. Vank\'{o}, R. Verbeni, and G. Monaco, 
\textit{Resonant inelastic hard x-ray scattering with diced analyzer crystals 
	and position-sensitive detectors},
Rev. Sci. Instr. \textbf{77}, 053102 (2006).

\bibitem{Moretti2013}
M. Moretti Sala, C. Henriquet, L. Simonelli, R. Verbeni, and G. Monaco, 
\textit{High energy-resolution set-up for Ir $L_3$ edge	RIXS experiments}, 
J. Elec. Spec. Rel. Phen. \textbf{188}, 150 (2013).

\bibitem{Minola15}
M. Minola, G. Dellea, H. Gretarsson, Y. Y. Peng, Y. Lu, J. Porras,
T. Loew, F. Yakhou, N. B. Brookes, Y. B. Huang, J. Pelliciari, T. Schmitt, G. Ghiringhelli, B. Keimer, L. Braicovich, and M. Le Tacon, 
Supplementary Information for \textit{Collective nature of spin excitations in superconducting cuprates probed by resonant inelastic x-ray scattering}, 
Phys. Rev. Lett. \textbf{114}, 217003 (2015).

\bibitem{Ament11}
L. J .P. Ament, M. van Veenendaal, T. P. Devereaux, J. P. Hill, and J. van den Brink,
\textit{Resonant inelastic x-ray scattering studies of elementary excitations},
Rev. Mod. Phys. \textbf{83}, 705 (2011).

\bibitem{Khan19}
N. Khan, D. Prishchenko, Y. Skourski, V. G. Mazurenko, and A. A. Tsirlin, 
\textit{Cubic symmetry and magnetic frustration on the fcc spin lattice in K$_2$IrCl$_6$}, 
Phys. Rev. B \textbf{99}, 144425 (2019).

\bibitem{ReigiPlessis20}
D. Reig-i-Plessis, T. A. Johnson, K. Lu, Q. Chen,
J. P. C. Ruff, M. H. Upton, T. J. Williams, S. Calder,
H. D. Zhou, J. P. Clancy, A. A. Aczel, and G. J. MacDougall, 
\textit{Structural, electronic, and magnetic properties of nearly ideal 
	$J_{\rm eff} = 1/2$ iridium halides},
Phys. Rev. Mater. \textbf{4}, 124407 (2020).

\bibitem{Goessling08}
A. G\"{o}ssling, R. Schmitz, H. Roth, M.W. Haverkort, T. Lorenz, J.A. Mydosh, 
E. M\"{u}ller-Hartmann, and M. Gr\"{u}ninger, 
\textit{Mott-Hubbard exciton in the optical conductivity of YTiO$_3$ and SmTiO$_3$}, 
Phys. Rev. B \textbf{78}, 075122 (2008). 

\bibitem{Reul12}
J. Reul, A. A. Nugroho, T. T. M. Palstra, and M. Gr\"{u}ninger, 
\textit{Probing orbital fluctuations in $R$VO$_3$ ($R$ = Y, Gd, or Ce) by ellipsometry}, 
Phys. Rev. B \textbf{86}, 125128 (2012).

\bibitem{Vergara22}
I. Vergara, M. Magnaterra, P. Warzanowski, J. Attig, S. Kunkem\"{o}ller, 
D.I. Khomskii, M. Braden, M. Hermanns, and M. Gr\"{u}ninger, 
\textit{Spin-orbit coupling and crystal-field splitting in Ti-doped Ca$_2$RuO$_4$ 
	studied by ellipsometry}, 
Phys. Rev. B \textbf{106}, 085103 (2022).

\bibitem{Henderson}
B. Henderson and G. F. Imbusch, 
{\it Optical spectroscopy of inorganic solids}, Oxford (1989).

\bibitem{Hitchman}
B. N. Figgis and M. A. Hitchman, 
\textit{Ligand Field Theory and its Applications} (Wiley) (1999).

\bibitem{Rueckamp05}
R. R\"uckamp, E. Benckiser, M. W. Haverkort, H. Roth, T. Lorenz, A. Freimuth, L. Jongen, 
A. M\"oller, G. Meyer, P. Reutler, B. B\"uchner, A. Revcolevschi, S.-W. Cheong, C. Sekar, 
G. Krabbes, and M. Gr\"{u}ninger, 
\textit{Optical study of orbital excitations in transition-metal oxides}, 
New J. Phys. \textbf{7}, 144 (2005).  

\bibitem{Benckiser08}
E. Benckiser, R. R\"{u}ckamp, T. M\"{o}ller, T. Taetz, A. M\"{o}ller, A. A. Nugroho, 
T. T. M. Palstra, G. S. Uhrig, and M. Gr\"{u}ninger, 
\textit{Collective orbital excitations in orbitally ordered YVO$_3$ and HoVO$_3$}, 
New J. Phys. \textbf{10}, 053027 (2008). 

\bibitem{Schmidt13}
M. Schmidt, Zhe Wang, Ch. Kant, F. Mayr, S. Toth, A. T. M. N. Islam, B. Lake, 
V. Tsurkan, A. Loidl, and J. Deisenhofer, 
\textit{Exciton-magnon transitions in the frustrated chromium antiferromagnets 
	CuCrO$_2$, $\alpha$-CaCr$_2$O$_4$, CdCr$_2$O$_4$, and ZnCr$_2$O$_4$}, 
Phys. Rev. B \textbf{87}, 224424 (2013).

\bibitem{Warzanowski20}
P. Warzanowski, N. Borgwardt, K. Hopfer, J. Attig, T. C. Koethe, P. Becker, V. Tsurkan, 
A. Loidl, M. Hermanns, P. H. M. van Loosdrecht, and M. Gr\"{u}ninger,
\textit{Multiple spin-orbit excitons and the electronic structure of $\alpha$-RuCl$_3$},
Phys. Rev. Res. \textbf{2}, 042007(R) (2020).

\bibitem{Hoggard81}
P. E. Hoggard,
\textit{Spin-Orbit Coupling in Tetragonal d$^3$ Systems},
Z. Naturforsch. \textbf{36a}, 1276 (1981).

\bibitem{Georges13}
A. Georges, Luca de'Medici, and J. Mravlje, 
\textit{Strong Correlations from Hund’s Coupling}, 
Ann. Rev. Condens.  Matt. Phys. \textbf{4}, 137 (2013).

\bibitem{Zhang17}
G. Zhang and E. Pavarini, 
\textit{Mott transition, spin-orbit effects, and magnetism in Ca$_2$RuO$_4$}, 
Phys. Rev. B \textbf{95}, 075145 (2017). 

\bibitem{Pross74}
L. Pross, K. R\"ossler, and H. J. Schenk, 
\textit{Optical studies on crystalline hexahalorhenates—I: Low temperature absorption 
	spectra of K$_2$[ReCl$_6$] single crystals}, 	
J. inorg. nucl. Chem. \textbf{36}, 317 (1974).

\bibitem{Yoo87}
R. K. Yoo, S. C. Lee, B. A. Kozikowski, and T. A. Keiderling, 
\textit{Intraconfigurational absorption spectroscopy of ReCl$_6^{2-}$ in various 
	$A_2M$Cl$_6$ host crystals},
Chem. Phys. \textbf{117}, 237 (1987).

\bibitem{Ballhausen}
C. F. Ballhausen, 
\textit{Introduction to Ligand Field Theory}, 
New York, McGraw-Hill (1962). 

\bibitem{Bettinelli88}
M. Bettinelli and C. D. Flint,
\textit{Magnon sidebands and cooperative absorptions in K$_2$ReCl$_6$ and Cs$_2$ReCl$_6$},
J. Phys. C: Solid State Phys. \textbf{21}, 5499 (1988).

\bibitem{Lorenzana95a}
J. Lorenzana and G. A. Sawatzky,
\textit{Phonon Assisted Multimagnon Optical Absorption and Long Lived Two-Magnon States in Undoped Lamellar Copper Oxides},
Phys. Rev. Lett. \textbf{74}, 1867 (1995)

\bibitem{Lorenzana95b}
J. Lorenzana and G. A. Sawatzky,
\textit{Theory of phonon-assisted
multimagnon optical absorption and bimagnon states in quantum antiferromagnets}, 
Phys. Rev. B \textbf{52}, 9576 (1995).

\bibitem{Windt01}
M. Windt, M. Gr\"uninger, T. Nunner, C. Knetter, K. P. Schmidt,
G. S. Uhrig, T. Kopp, A. Freimuth, U. Ammerahl, B. B\"uchner, and A. Revcolevschi, 
\textit{Observation of Two-Magnon Bound States in the Two-Leg Ladders of 
	(Ca,La)$_{14}$Cu$_{24}$O$_{41}$}, 
Phys. Rev. Lett. \textbf{87}, 127002 (2001).

\bibitem{Kamimura60}
H. Kamimura, S. Koide, H. Sekiyama, and S. Sugano, 
\textit{Magnetic Properties of the Pd and Pt Group Transition Metal Complexes}, 
J. Phys. Soc. Jpn.  \textbf{15}, 1264 (1960).

\bibitem{Figgis61}
B. N. Figgis, J. Lewis,and F. E. Mabbs, 
\textit{The Magnetic Properties of Some $d^3$-Complexes}, 
J. Chem. Soc. 3138 (1961).

\bibitem{Szymczak80}
H. Szymczak, W. Wardzy\'nski, and A. Pajaczkowska, 
\textit{Optical spectrum of antiferromagnetic spinels ZnCr$_2$O$_4$}, 
J. Magn. Magn. Mater. \textbf{15-18}, 841 (1980).

\bibitem{Eremin11}
M. V. Eremin and M. A. Fayzullin,  
\textit{Possible mechanisms of magnon sidebands formation in transition metal compounds}, 
J. Phys.: Conf. Ser. \textbf{324}, 012022 (2011).

\end{thebibliography}
\end{document}